\documentclass[aps,prab,reprint,superscriptaddress,showpacs,showkeys,letter]{revtex4-1}
\usepackage{graphicx}

\usepackage{dcolumn}
\usepackage{bm}
\usepackage{amsmath}
\usepackage[caption=false]{subfig}
\usepackage[none]{hyphenat}
\usepackage{color}
\usepackage{mathrsfs}
\usepackage{epstopdf}
\usepackage{amssymb}
\usepackage{mathrsfs}
\usepackage{url}

\begin{document}


\title{Studies of the beam-ion instability and its mitigation with feedback system}

\author{Chao Li}
\email{lichao@ihep.ac.cn}
\author{Saike Tian}
\author{Na Wang}
\author{Haisheng Xu}
\affiliation{Key Laboratory of Particle Acceleration Physics and Technology, Institute of High Energy Physics, Chinese Academy of Sciences, 19(B) Yuquan Road, Beijing 100049, China}

\begin{abstract}
The beam-ion interaction is a potential limitation of beam performance in electron accelerators, especially where the beam emittance is of a great concern in future ultra-low emittance light source. ``Conventionally", the beam instability due to beam-ion interaction is attributed to two types of effects:  ion trapping effect and  fast ion effect, which emphasize the beam-ion dynamics in different time scales. Whereas, in accelerators, the beam suffers from a mixture of ion trapping effect and fast ion effect, leading to a more complicated process and requiring a self-consistent treatment. To evaluate the beam characteristics, as emittance growth under the influence from beam-ion effect, a new numerical simulation code based on the ``quasi-strong-strong" model has been developed, including modules of ionization, beam-ion interaction, synchrotron radiation damping, quantum excitation, bunch-by-bunch feedback, etc. In the study, we do not regularly distinguish the ion trapping effect and the fast ion effect, but treat beam-ion interaction more generally and consistently. The lattice of High Energy Photon Source, a diffraction limit ring under construction in Beijing, is used as an example to show the beam-ion effect. It is found that in this low emittance ring, the beam-ion instability is not a dominant mechanism in operation mode with a  high beam current, but seriously occurs in a lower beam current region. When the  beam-ion  instability were significantly driven and can not be damped by the synchrotron radiation damping, the simulations show the bunch-by-bunch feedback system based on the Finite Impulse Response  filter technique can be adopted to mitigate it effectively.
\end{abstract}
\maketitle
\clearpage

\section{Introduction}  

The beam-ion interaction, a two-stream effect coupled by the nonlinear Coulomb force, may pose a risk to the operation of future electron accelerators with beams of high intensity and ultra-low emittances.  The effect of beam-ion interaction has been observed in many existing accelerator machines such as ALS \cite{1, 2}, PLS \cite{3}, SSRF \cite{4} CESR-TA \cite{5}, SOLEIL \cite{6}, etc. The accumulated ions, derived from ionization between electron particle and residual gas molecules, interact with the electron beam particles resonantly, causing  coherent and incoherent electron beam deformation, such as beam centroid oscillation, beam rms emittance growth, rms beam sizes increasement, energy spread blow up, and even a possible beam loss.  

In the previous studies,  the beam-ion effect \cite{7,8,9} is divided into two circumstances known as ion trapping effect and fast ion effect. In the study of ion trapping effect, one of the key conclusions is the ion trapping condition \cite{8}, 
\begin{eqnarray}
\label{eq1.1}
A>A_{th} = \frac{N_b r_p \Delta T_b c}{2 \sigma_{x,y} (\sigma_{x} +\sigma_{y}) }, 
\end{eqnarray}
where $N_b$ is the bunch population, $r_p$ is the classical proton radius, $\Delta T_b$ is the bunch spacing, $c$ is the speed of light, $\sigma_x$ and $\sigma_y$ are the rms beam size. Ions with mass number lower than $A_{th}$ will be over-focused by the bunched beam particles and hardly disturb the beam performance. Ions with a mass larger than the critical value $A_{th}$ will be trapped transversely in the space charge potential well of the electron beam and impact the electron beam particles turn by turn leading to a beam performance deterioration, or even beam losses. It is noteworthy that the ion trapping condition is based on the linear space charge and even beam filling pattern assumptions; moreover, the critical mass $A_{th}$ varies along the accelerator since the betatron function varies, which indicates the ions trapping sections are localized actually. To simplify the theoretical analysis and approaches adopted in numerical simulations for ion trapping study,  it is usually assumed that the accumulated ions are saturated to an equilibrium state over thousands turns -- a steady state sense. In this equilibrium state frame, a beam-ion eigen-system can be established and used to predict the coupling resonance conditions, which is a typical methodology in two stream instability studies \cite{10}. Results obtained from ion trapping  are usually used as guidances for vacuum selection and beam intensity up-limits prediction. 

Correspondingly, the fast ion effect focuses on the transient effect in one turn. The ions generated in the first turn are cleaned and will not disturb the beam performance in the second turn. In the time scale of one turn, the ions generated by the leading bunches oscillate transversely and resonantly disturb the motions of the subsequent bunches -- coupled bunch instability in a transient sense. Then the ions are assumed to be cleaned over one turn. Generally, the fast ion effect disturbs the beam performance both in linacs and storage rings in a spontaneous manner. 

In general, the transient net ions accumulated is a compromise of ions generation rate due to ionization and ion loss rate due to momentum kicks from bunched beam passed by. Clearly, the ion trapping effect and fast ion effect are both derived from the beam-ion Coulomb interaction, but emphasize the dynamics process in different time scales.  In reality, these approximations adopted in ion trapping effect and fast ion effect are always violated by constraints such as uneven beam filling pattern, residual ions accumulated turn by turn, etc, which lead to the  theoretical task impossible to complete. In reality, ion trapping and fast ion effect could take place simultaneously.   In this paper, we do not distinguish the ion-trapping and fast ion effects and treat the beam-ion motions consistently. To evaluate the beam performance, such as beam emittance evolution, a new numerical code based on the ``quasi-strong-strong" approach is developed \cite{11}, in which electron particles and ions are both represented by multi-macroparticles. Moreover, modules of ionization, beam-ion interaction, synchrotron radiation damping, quantum excitation and bunch-by-bunch feedback are also established. As an example, High Energy Photon Source (HEPS) \cite{12} lattice is adopted to show the beam-ion effect in detail. It will be shown that in this ultra-low emittance ring, the beam-ion interaction significantly influences the beam performance only when ions can be extensively accumulated in certain current region. If the beam current in the operation is high enough, the ions will be over-focused, get lost in gaps between neighbouring bunches, and hardly influence the beam.

To suppress the beam-ion instability, the bunch-by-bunch feedback system is applied in this paper. Generally, the methodologies to mitigate the beam-ion effect are: (1) adjust the beam filling pattern by including empty buckets long enough in the bunch train; (2) get rid of ions with certain accelerator elements, such as cleaning electrodes where ions can be collected; (3) cure the beam-ion instability by introducing a feedback system before it grows \cite{13}. The first approach can extensively reduce the number of accumulated ions. With sufficient large empty gaps, the trapped ions would drift to large amplitudes, where they may get lost on the pipe or form a diffuse ion halo that hardly influences the beam. However, this approach is a partial solution since the disturbed bunch can not erase the memory itself. The beam deformation by the beam-ion interaction will accumulate and the influence will be shown finally, unless the synchrotron radiation damping is pretty strong. The second and the third approaches both require extra  hardware, which brings in new sources of lattice impedance. However, the bunch-by-bunch feedback system is a versatile system \cite{14}. It also can be used to suppress the beam instabilities lie in impedance.

This paper is organized as follows, in section II, the physical process and models of beam-ion interaction will be discussed briefly. The logic flow and  basic approaches used in the code will be given. In section III, both the ``weak-strong" and ``quasi-strong-strong" simulations of the beam-ion effect in HEPS  will be discussed in detail.  The bunch-by-bunch feedback will be discussed in Section V. By introducing an appropriate feedback system, the beam-ion instability can be effectively suppressed. The discussion and conclusion are given in Section IV.  

\section{Physical model and the logical flow in the code developed.} 
Ignoring the ions generated from the synchrotron radiation, which are usually far outside the beam and equally distributed between the beam and chamber wall, denoting $P$ and $T$ as the vacuum pressure and temperature, the molecules density $n$ in the accelerator can be obtained from the general gas equation, 
\begin{eqnarray}
\label{eq2.1}
P N_A= nRT,
\end{eqnarray}
where $R$ and $N_A$ are the ideal gas constant and the Avogadro number. Denote $\sum$ as the ionization cross-section, $N_b$ as the number of electron particles passing by, the number of ionization ions per unit length is 
\begin{eqnarray}
\label{e2.2}
\lambda = \sum n N_b.
\end{eqnarray}

For simplicity, the interaction between ions and beam is assumed taking place at lumped interaction locations, and the ions are assumed not to move longitudinally. 
It is noteworthy that, when beam passes through interaction points bunch by bunch, new ions will be repeatedly generated with a minimum time interval $T_{rf}$, which is the period of fundamental radio frequency indicating the spacing between adjacent bunches. The ions generated  are randomly distributed  in the same range as the size of the electron bunch passing by. Meanwhile, the accumulated ions are kicked by the passing bunched electron particles and then drift freely until next electron bunch comes. Some of the ions might get lost on the pipe. Due to the ions generation and loss mechanisms,  a dynamical quasi-equilibrium ion distribution can be foreseen finally. The motion equations of the $i$th accumulated ion $\vec X_i$ and the $k$th electron particle in the $j$th bunch $\vec x_{k;j}$ can be expressed as
\begin{eqnarray}
\label{eq2.3}
\frac{d^2\vec X_i}{dt^2} +K_i(s) \vec x_{k;j} + \sum_{k=0}^{N_j} \vec F_C(\vec X_i - \vec x_{k;j}) =0 \nonumber  \\
\frac{d^2\vec x_{k;j}}{ds^2} + K_e(s) \vec x_{k;j} +\sum_{i=0}^{N_i} \vec F_C(\vec x_{k;j} - \vec X_i ) =0,
\end{eqnarray}
where $\vec F_C$ is the Coulomb force between the ions and electron particles, $K_i(s)$ and $ K_e(s)$ represent the lattice focusing strength on ion and beam particle.

\begin{figure}[t]
\centering
\includegraphics[width=1\linewidth]{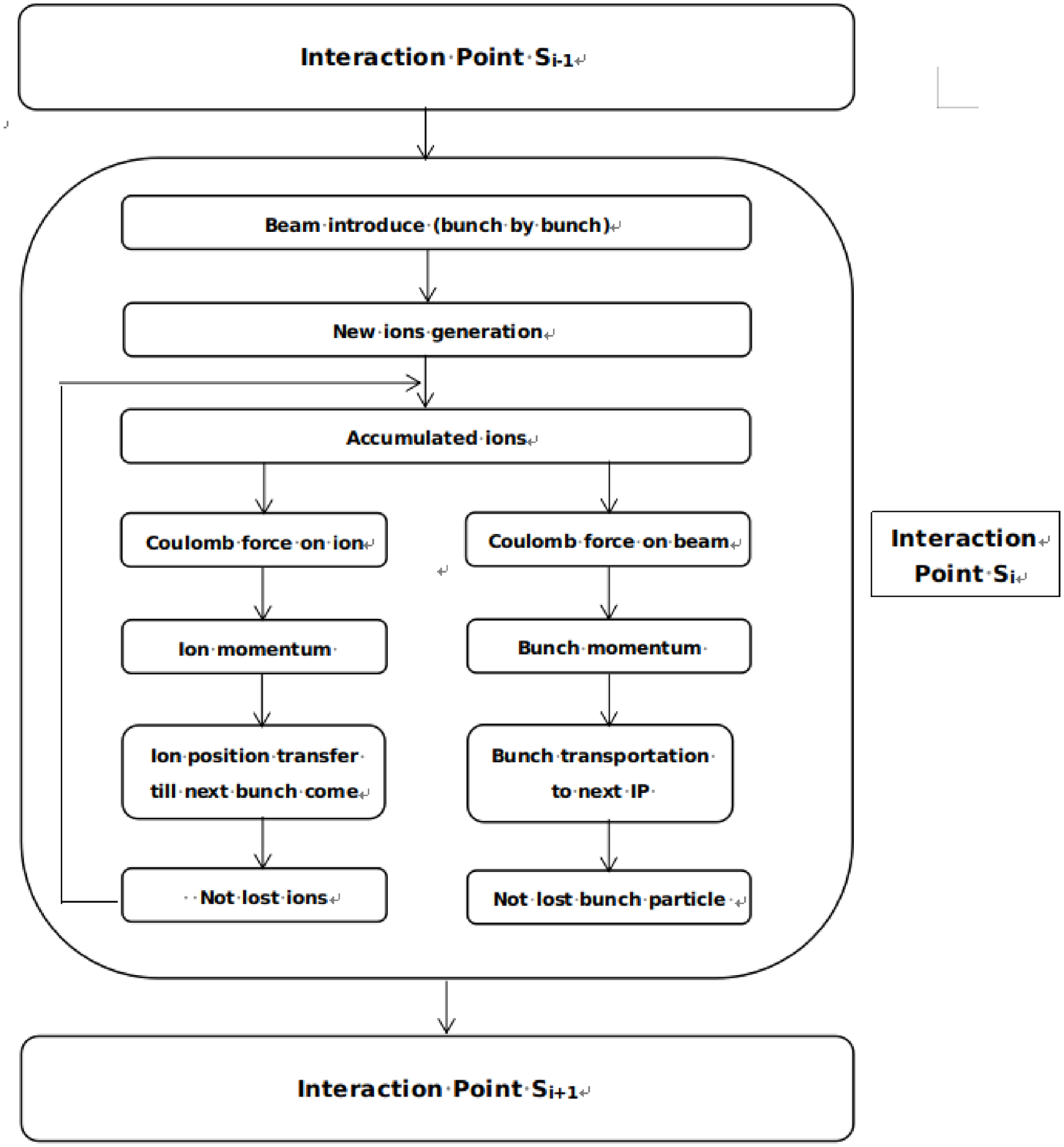}\\
\caption{Logic flow of the beam-ion interaction in simulation at the interaction points. }
\label{fig:fig1}
\end{figure}

Eq.~\ref{eq2.3} describes the beam-ion dynamics in a general sense but almost impossible to  be applied directly in theoretical analysis.  To launch that, several approaches can be adopted to simplify the analytical study, as
\begin{itemize}
\item[1. ] the smooth approximation to get a time-independent focusing lattice; 
\item[2. ] the equilibrium state assumption of the accumulated ions -- constant ion charge; 
\item[3. ] the linear space charge approximation.
\end{itemize}  
More detailed information can be found in Ref.~\cite{7, 8}. 

Here, we briefly introduce the numerical approaches used in the code developed \cite{11}. Fig.~\ref{fig:fig1} shows the logic flow of the simulation process at the interaction points. The continuous interaction between ions and beam particles is limited to  lumped interaction locations. Arbitrary number of interaction points with various local lattice functions, Twiss parameters, temperatures and gas pressures can be specified in the code. In the simulation model, both the ions and  electron bunches are represented by macro-particles. The electron bunch with  $N_b$  particles is assumed to  follow a rigid Gaussian distribution in real space, which is rather reasonable. At the interaction point $s_i$, the 2D Bassetti-Erskine formula \cite{15} is used to get electron field generated by the bunched electron particles, 
\begin{eqnarray}
\label{eq2.5}
E_{C,y}( \vec  x ) + I E_{C, x}(\vec x) &=& \frac{n_b}{2\epsilon_0\sqrt{2\pi (\sigma_x^2-\sigma_y^2)}} 
\{w(\frac{x+Iy}{\sqrt{2(\sigma_x^2-\sigma_y^2)}}) \nonumber  \\
&-&\exp(-\frac{x^2}{2\sigma_x^2}-\frac{y^2}{2\sigma_y^2}) \nonumber  \\
 &+& w(\frac{x\frac{\sigma_y}{\sigma_x}+Iy\frac{\sigma_x}{\sigma_y}}{\sqrt{2 (\sigma_x^2-\sigma_y^2)}}) \} \delta(s_i), 
\end{eqnarray}
where $n_b$ is the line density of the electron beam, $w(z)$ is the complex error function, $\sigma_x$ and $\sigma_y$ are the beam rms size in horizontal and vertical direction, $x$ and $y$ are the distance from ions to the bunch centroid, $I$ is the complex unit. Substituting Eq.~\ref{eq2.5} into Eq.~\ref{eq2.3}, the explicit momentum change of ions at the interaction point is 
\begin{eqnarray}
\label{eq2.6}
\Delta p_{i,y} + I \Delta p_{i,x} &=& \frac{2 n_br_em_ec}{\gamma_e} (E_{C,y} + I E_{C, x}), 
\end{eqnarray}
where $r_e$ is the classical electron radius, $m_e$ is the electron mass, $c$ is the speed of light, $\gamma_e$ is the relativistic factor of electron beam. Since the ions are much heavier than the electron, the lattice focusing $K_i(s)$ can be ignored. Integrating Eq.~\ref{eq2.6} along the length of adjacent electron bunches, the accumulated ion momentum change induced by the electron bunch passed by can be obtained.

As to the space charge potential well generated by the ions, since the ions distribution is usually not a Gaussian type and the ion particles almost occupy the whole pipe, in principle the Bassetti-Erskine formula is not suitable anymore. A self-consistent particle-in-cell (PIC)  \cite{16} solver or ion density profile fitting \cite{17} is needed to ensure a better resolution. In our code, a  compromise approach is applied. The ions distribution is truncated at 10 rms bunch size. The rms and centroid information of the truncated ion distribution are substituted in the Bassetti-Erskine formula to get the Coulomb potential. Although this approach is not as self-consistent as  PIC, it still  can show the main features of the bunched beam and explore this complex coupled dynamics in a reasonable computing time. The vacuum chamber aperture is used as the ions and electron particle loss criteria; Only the survived ions and  electron particles are kept for further calculation. When one electron bunch particles passes by, the transverse momentum and position of the accumulated ions are updated according to the time interval until the next bunch comes. As to the bunched electron particles, after the momentum kicks induced by the accumulated ions, they are transferred to the next interaction point by applying the linear transport matrix.

With a monotonically increasing of the bunched beam size due to beam-ion interaction, a saturated accumulated ions with a sharp center density profile can be foreseen \cite{13}. In our code, both the electron bunched beam and ions are represented by multi-particles -- ``quasi-strong-strong" model. Compared with the linear space charge assumption or ``weak-strong" model, where the growth rate of the beam-ion instability is overestimated, the ``quasi-strong-strong" includes ion and beam transverse oscillation frequency spreads spontaneously that  eases the beam-ion instability through Landau damping. However, it is still noteworthy that there are gaps between the Gaussian profile assumption and the real beam (or ion) distributions. To get a more self-consistent process and better accuracy, a general Poisson solver is required in further study .

Summing up, the characteristics of code are: 
\begin{enumerate}
\item settings of arbitrary ion species, arbitrary number of beam-ion interaction points where the local gas pressure and temperature are specified; 
\item including synchrotron radiation damping and quantum excitation;
\item including bunch-by-bunch feedback modules;
\item ``quasi-strong-strong" model;
\item Gaussian distribution assumption in the beam-ion space charge calculation;
\end{enumerate}

\section{Simulation study of beam-ion interaction in HEPS without feedback.} 
\begin{table}[]
\caption{HEPS Lattice Parameters}
\begin{tabular}{l l  }
\hline
Parameters &  Values          \\
\hline
Energy      & 6 GeV              \\
\hline
Circumference     & 1360.4 m     \\
\hline
Nominal emittance & 34.2 pm     \\
\hline 
Working points   &  114.14/106.23  \\
\hline
Number of super-periods & 24  \\
\hline
Average betatron function &  4.5/8.1 m  \\
\hline
Number of RF bucket &  756\\
\hline
Beam current & 200 mA\\
\hline
SR damping time  (x/y) & 2386/4536 turns \\
\hline
rms beam size  (x/y) & 12.4/5.26 $\mu$m \\
\hline 
Ion species   & CO \\
\hline
Gas pressure   & 1 nTorr \\
\hline
Gas Temperature   & 300 K \\
\hline
\hline
\end{tabular}
\label{Tab:table1}
\end{table}

HEPS is a 1.3 km ultra-low emittance electron storage photon source being built in Beijing, China. The main parameters of HEPS lattice are listed in Tab. I. Carbon monoxide (CO)  is assumed as the leaked gas with pressure 1 nTorr and temperature 300 K. At the beam operation stage with 200 mA beam current, the critical mass number of ion trapping condition is $A_{th}\approx 120$ \footnote{The average betatron function is used to estimate the ion trapping condition.}. In the following study, the total electron beam current 10 mA ($A_{th}\approx 10$) is adopted to evaluate the beam-ion effect, since it gives the most serious beam-ion effect which will be explained below. To save computing time, one beam-ion interaction point is set per turn. The beam filling pattern is one continuous bunch train following 76 empty bunch gaps. The synchrotron radiation damping and quantum  excitation  \cite{18} are both taken into account.

In this section, results from ``weak-strong" and ``quasi-strong-strong" model  both will be   discussed in detail. In the ``quasi-strong-strong” model, the beam and ion density variations are intrinsic included, leading to ion and beam transverse oscillation frequency spreads. These frequency spreads are supposed to ease the beam-ion instability through Landau damping. Another necessity of the ``quasi-strong-strong" simulation study is the beam emittance growth evaluation, which is one of the key challenges in ultra-low emittance rings.

\begin{figure*}[htb!]
    \centering
    \subfloat[\label{sfig:WSFillingPattern0_1}]{
        \includegraphics[width=.45\linewidth]{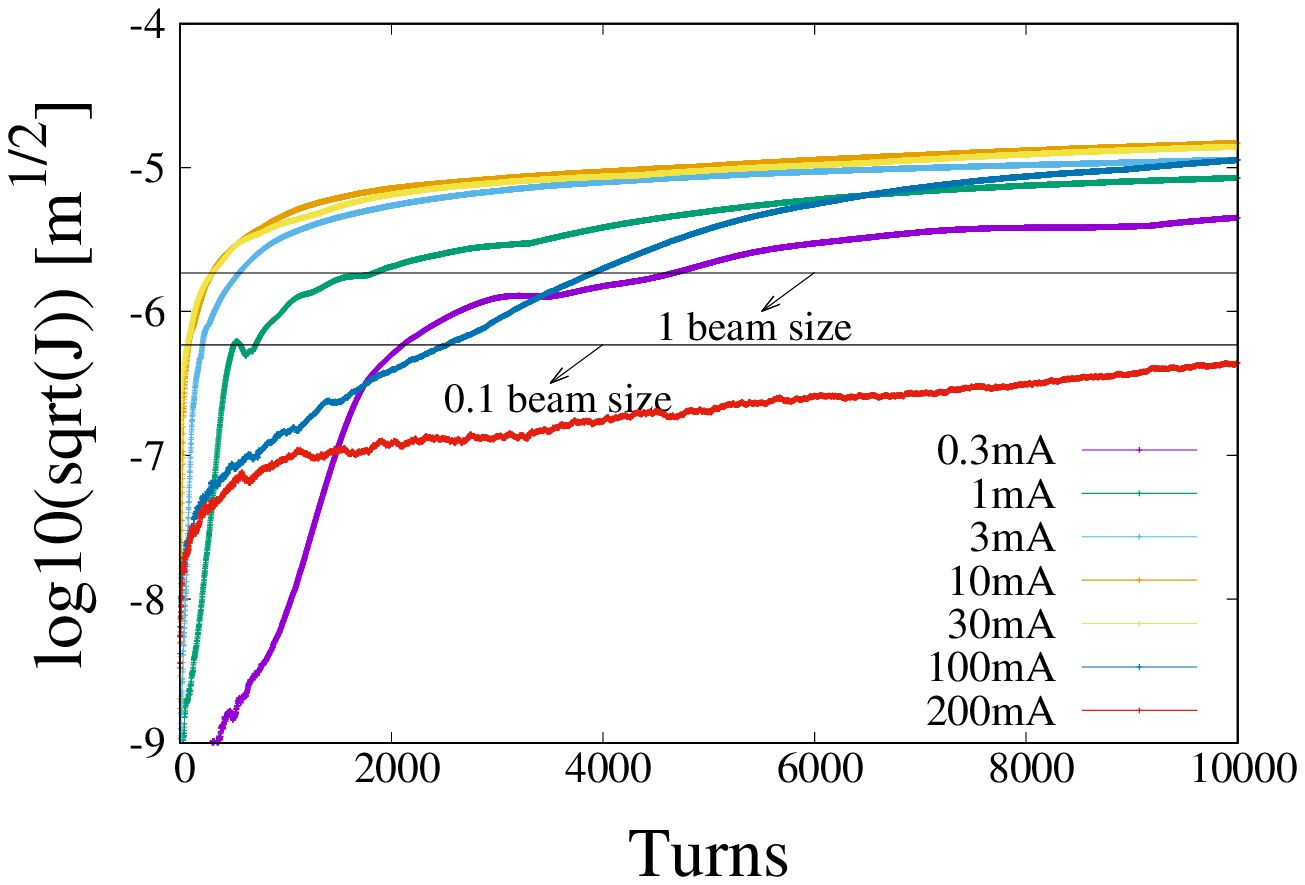}
    }
    \subfloat[\label{sfig:WSFillingPattern0_2}]{
        \includegraphics[width=.45\linewidth]{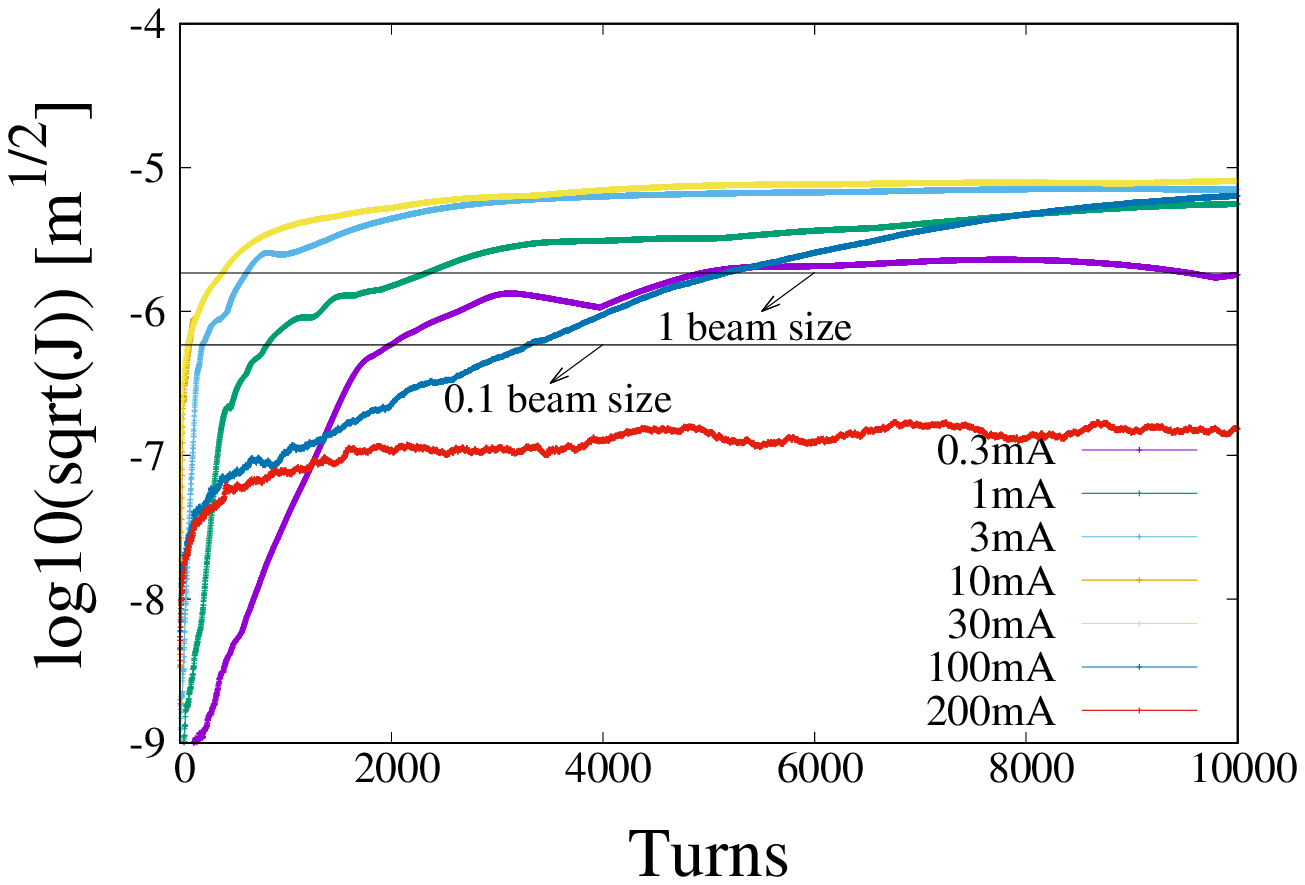}
    }
    \caption{The  maximum bunch action at the interaction point as function of tracking turns in ``weak-strong" simulation without (left) and with (right) synchrotron radiation damping. In each subfigure, there are 7 curves related to beam current settings 0.3 mA, 1 mA, 3 mA, 10 mA, 30 mA, 100 mA and 200 mA.  The synchrotron radiation damping is not able to mitigate the beam-ion instability.}
    \label{fig:WSFillingPattern0}
\end{figure*}

\subsection{``Weak-Strong'' Simulation}
In the ``weak-strong" model, the electron bunch is represented by a rigid Gaussian distribution. In the code, setting one macro electron particle per bunch, the simulation automatically degenerates to the ``weak-strong" case. Smaller vertical beam emittance leads to more serious beam deformation in vertical space. Following the approaches used in previous researches \cite{9,19}, the maximum bunch dipole moment is recorded turn by turn. The vertical amplitude of the bunch centroid oscillation is half of the Courant-Synder invariant, which is give by 
\begin{eqnarray}\label{eq3.01}
J_y = \frac{1}{2}(\frac{1+\alpha_y^2}{\beta_y}y^2+2\alpha_y  y y' + \beta_y y'^2) 
\end{eqnarray}
where $\alpha_y$ $\beta_y$ are the local Twiss parameters at the interaction point.

\begin{figure*}[htb!]
    \centering
    \subfloat[\label{sfig:Syn_10mA_Mode_and_Oscillation_1}]{
        \includegraphics[width=.45\linewidth]{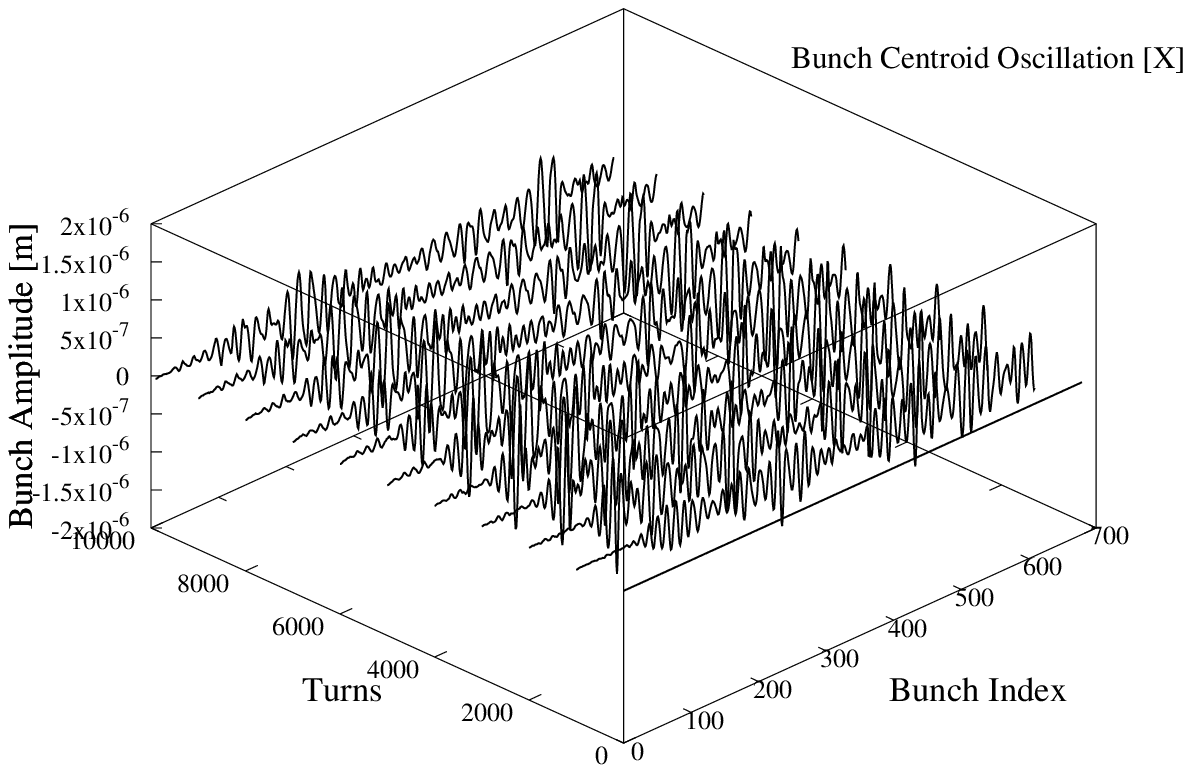}
    }
    \subfloat[\label{sfig:Syn_10mA_Mode_and_Oscillation_2}]{
        \includegraphics[width=.45\linewidth]{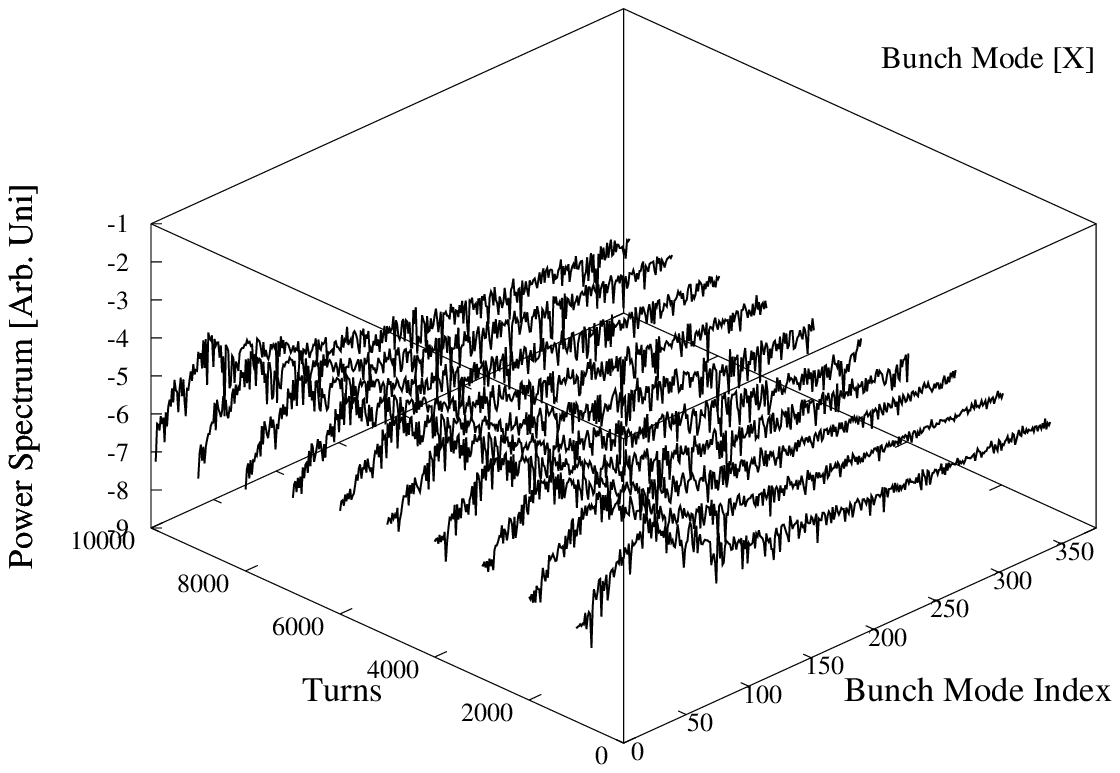}
    }
    
    \subfloat[\label{sfig:Syn_10mA_Mode_and_Oscillation_3}]{
        \includegraphics[width=.45\linewidth]{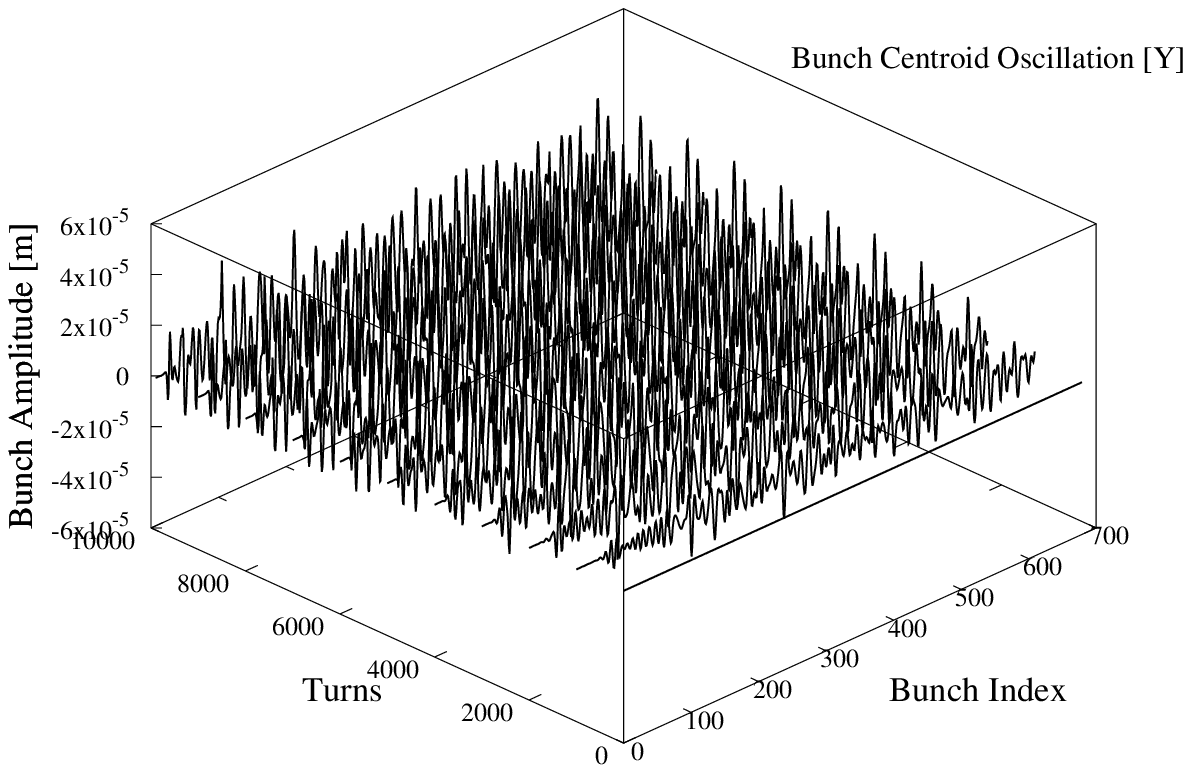}
    }
    \subfloat[\label{sfig:Syn_10mA_Mode_and_Oscillation_4}]{
        \includegraphics[width=.45\linewidth]{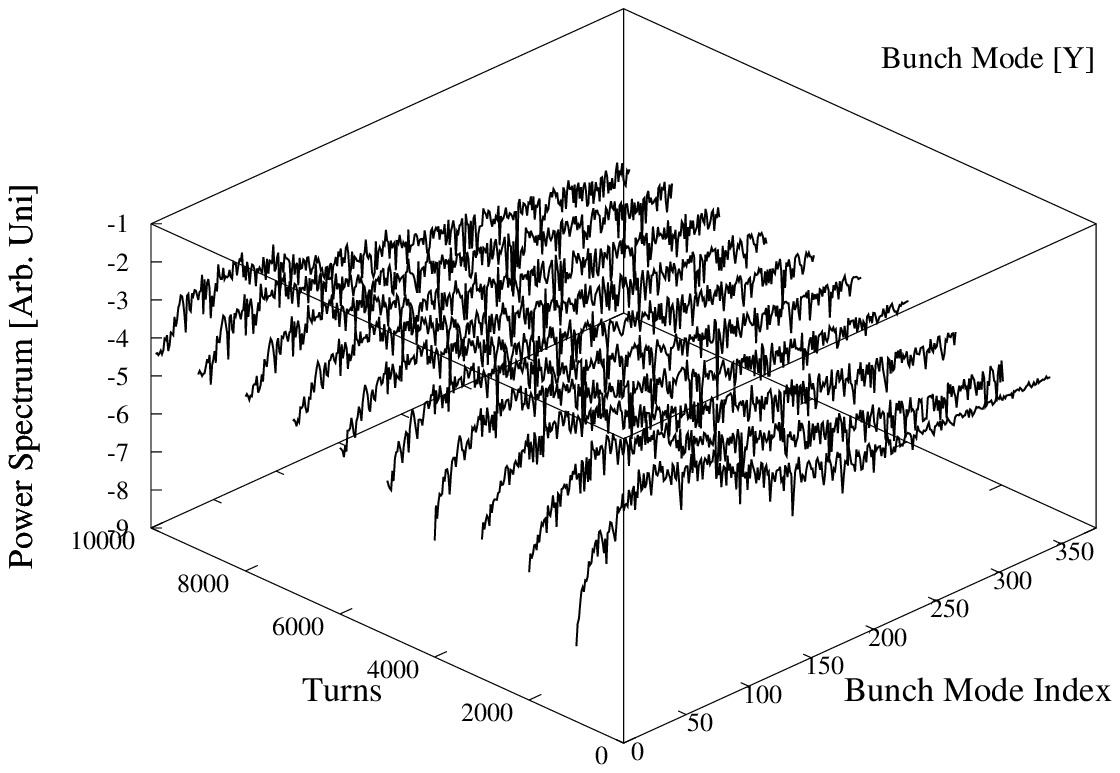}
    }
    \caption{Beam bunches oscillations in horizontal (top) and vertical (bottom) and its frequency spectrum  as function of passing turns. Clear coupled bunch motions can be identified. The simulation results are given by ``weak-strong" model with the synchrotron radiation damping.}
    \label{fig:Syn_10mA_Mode_and_Oscillation}
\end{figure*}

Fig.~\ref{fig:WSFillingPattern0} shows the maximum bunch action $\sqrt{J_y}$ as function of tracking turns in ``weak-strong" simulation. The sub-figures at left and right correspond to simulation without and with synchrotron radiation damping. In each sub-figure, there are 7 curves related to beam currents: 0.3 mA, 1 mA, 3 mA, 10 mA, 30 mA, 100 mA and 200 mA. For each beam current, the maximum bunch action $\sqrt{J_y}$ performs a sustained increasing due to the beam-ion interaction and finally arrive to a ``constant" value after 10 thousands turns evolution.  The final beam action $\sqrt{J_y}$ increases with beam current firstly and reaches a maximum value (10 mA) and then decreases. This is because of the net accumulated ion number is a nonlinear  function verse beam current. Fig.~\ref{fig:WSFillingPattern0} also depicts that the beam is still significantly impacted by the ions till 100 mA. Up to 200 mA, the bunched beam is hardly impacted and the centroid oscillation is smaller than 0.1 rms beam size, which is because the trapping condition can not hold anymore and the ions are over-focused and got lost in the empty gaps between the neighbouring bunches. Comparing the results  with and without synchrotron radiation damping, the final bunch actions $\sqrt{J_y}$ become smaller and get into a equilibrium state sooner when the synchrotron radiation damping is turned on.

It is noteworthy that the spontaneous synchrotron radiation damping in HEPS can not stabilize the beam-ion instability except the ions are over-focused in 200 mA case. In the following discussion, if not stated, 10 mA beam current is chosen as a default case to show the characteristics of the beam-ion interaction. Fig.~\ref{fig:Syn_10mA_Mode_and_Oscillation} illustrates the bunch centroid oscillation and its coupled mode spectrum  when the synchrotron radiation damping is turned on. The oscillations of different bunches show a clear coupled ``head-tail" style motion as expected. With the turn number increasing, more and more bunches at the tail part start to oscillate which is one of the typical characteristics of the beam-ion instability. The maximum unstable mode index is around 50 (70) in $x$ ($y$) direction.

\begin{figure*}[htb!]
    \centering
    \subfloat[\label{sfig:WSIONAccumu_1}]{
        \includegraphics[width=.45\linewidth]{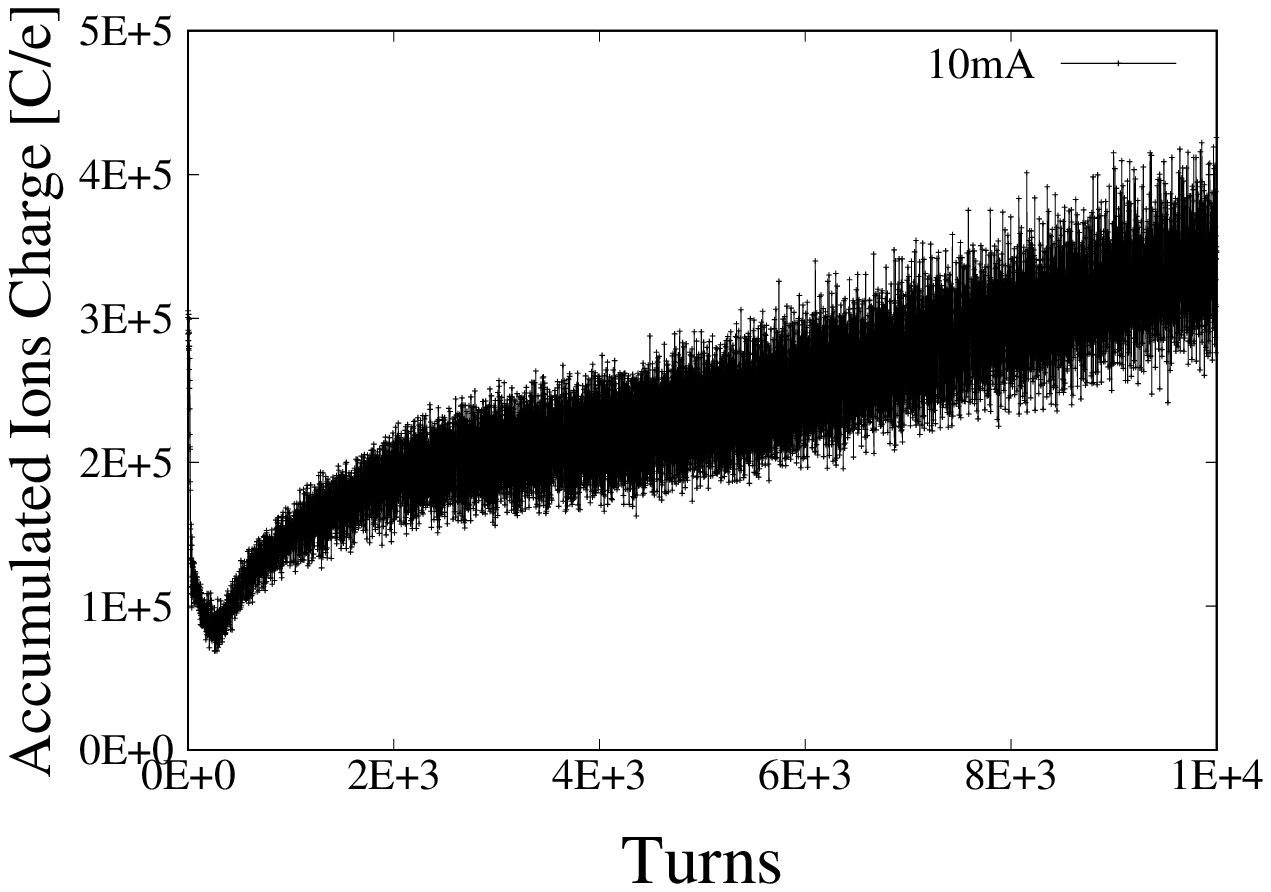}
    }
    \subfloat[\label{sfig:WSIONAccumu_2}]{
        \includegraphics[width=.45\linewidth]{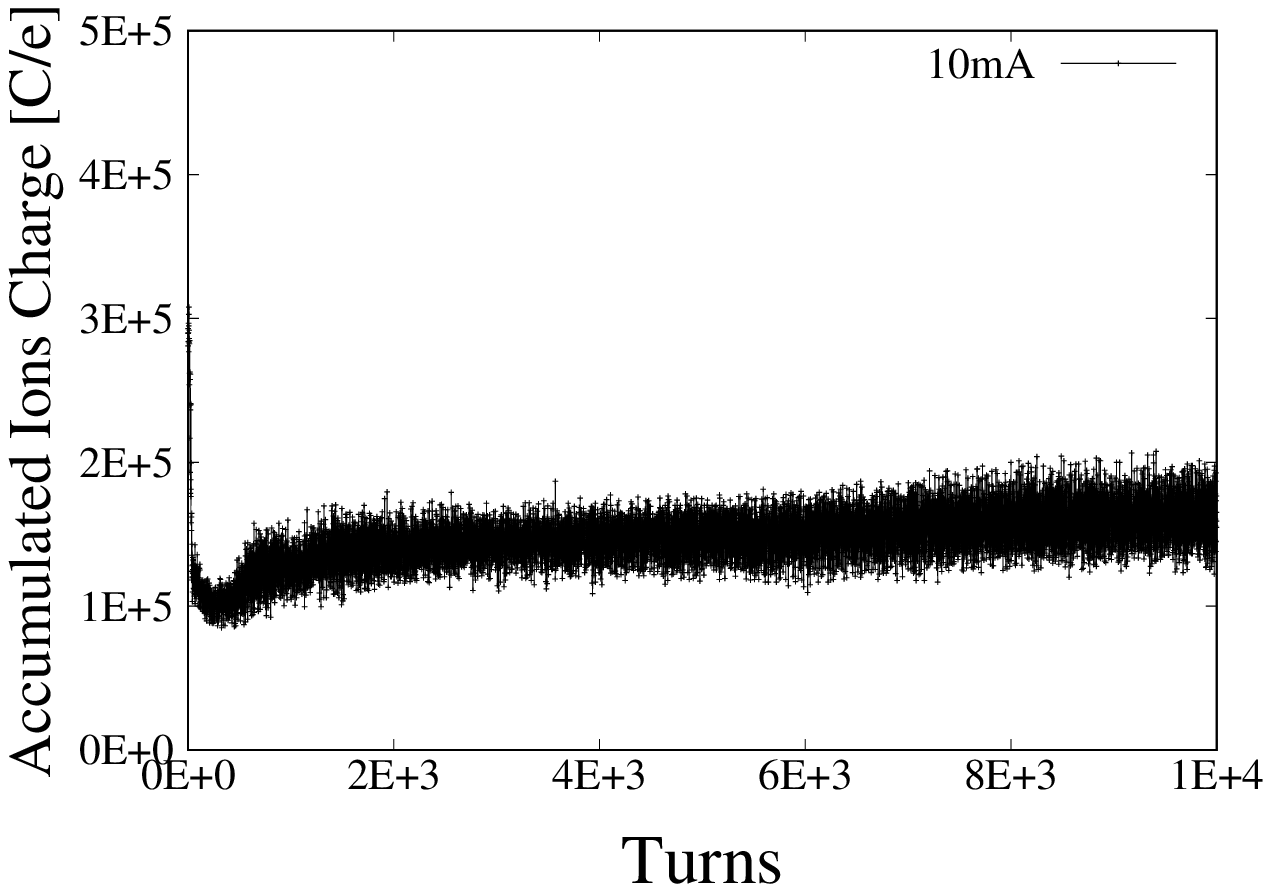}
    }
    \caption{The total ions charge accumulated at the interaction point as function of tracking turns in ``weak-strong" simulation without (left) and with (right) synchrotron radiation damping. }
    \label{fig:WSIONAccumu}
\end{figure*}

\begin{figure*}[htb!]
    \centering
    \subfloat[\label{sfig:FigIonDenProfile_0}]{
        \includegraphics[width=.3\linewidth]{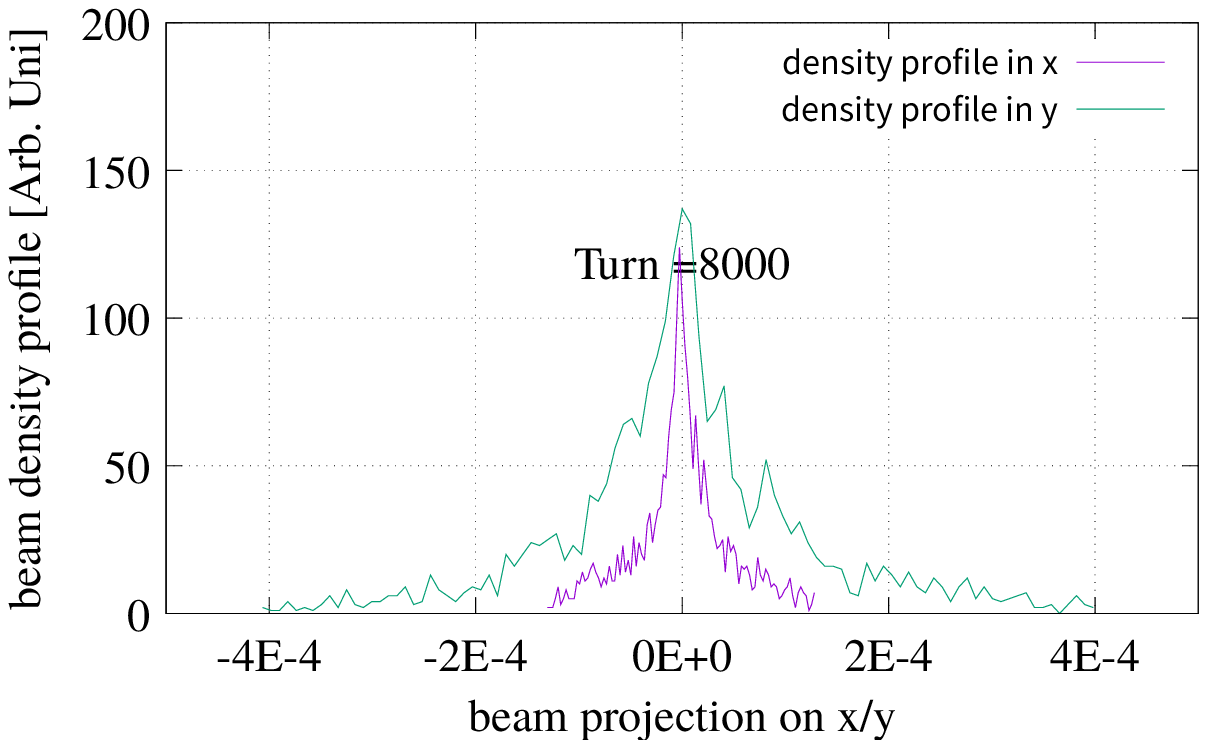}
    }
    \subfloat[\label{sfig:FigIonDenProfile_4}]{
        \includegraphics[width=.3\linewidth]{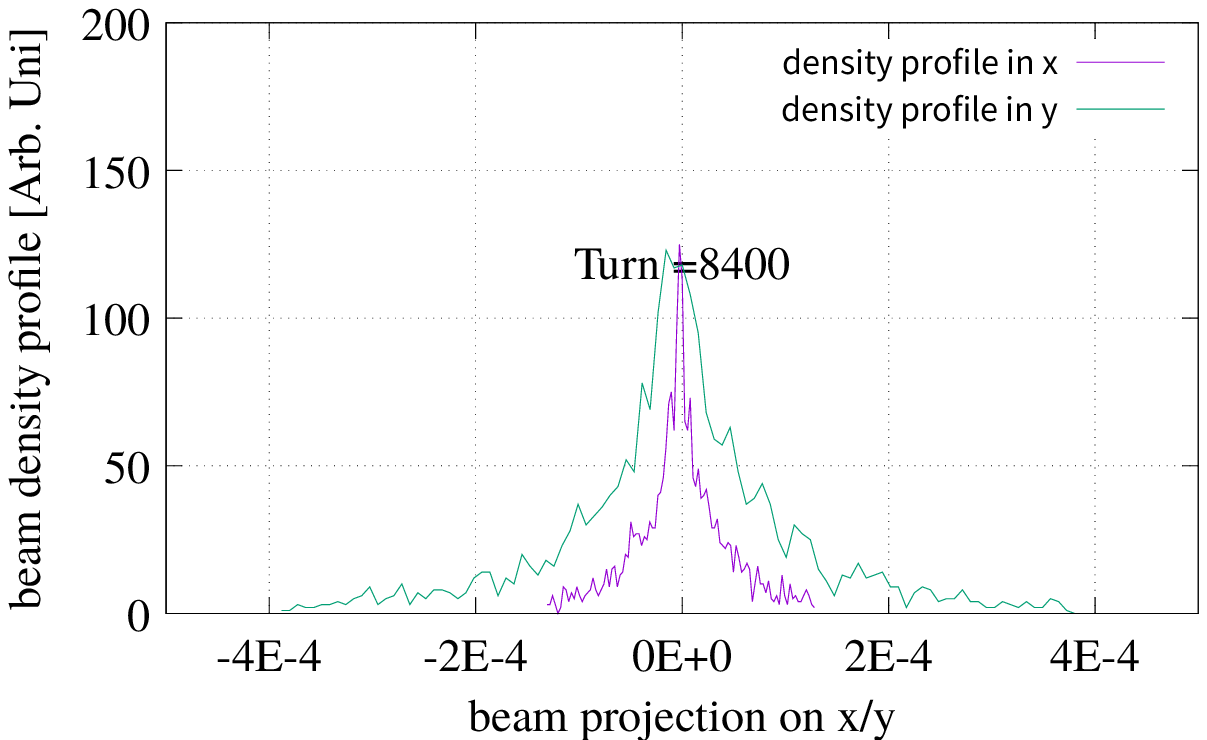}
    }
    \subfloat[\label{sfig:FigIonDenProfile_9}]{
        \includegraphics[width=.3\linewidth]{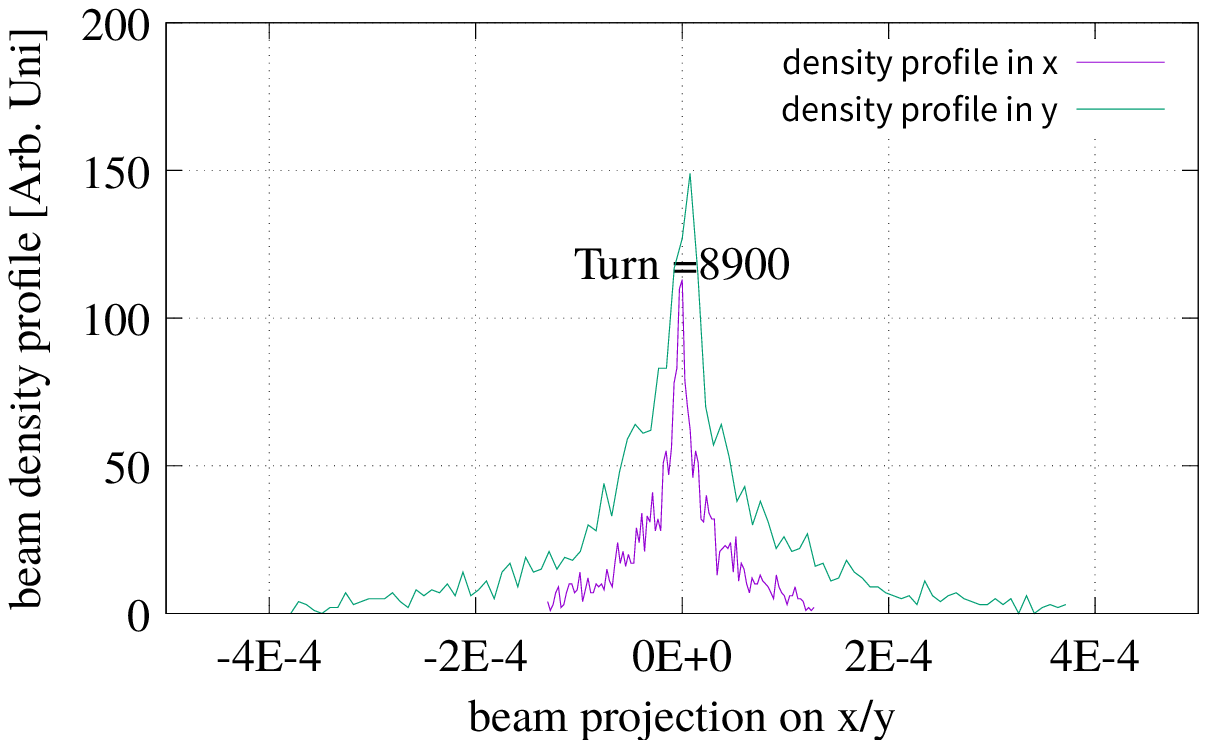}
    }
    \caption{Ion density profile sampled at the 8000th, 8400th, 8900th turns (from left to right) . The ion density profiles deviate from the Gaussian type and  form a higher density peak in the center. The density profile in the vertical direction shows an oscillation due to the beam-ion interaction. The simulation is launched with ``weak-strong" model and takes the synchrotron radiation damping into account.  }
    \label{fig:FigIonDenProfile}
\end{figure*}

In simulations, the ion particle is marked as a lost one either it beyonds  10 times effective rms beam size or beyonds the pipe aperture. Fig.~\ref{fig:WSIONAccumu} shows the total accumulated ion charge as function of tracking turns in the ``weak-strong" model. Similar to Fig.~\ref{fig:WSFillingPattern0}, the figures at left and right correspond to results without and with synchrotron  radiation damping. Dynamically, the accumulated ion charge is a compromise between ions generation rate and ions loss rate. The ion loss rate lies in the ion's accumulated  transverse momentum obtained from the electron bunch. With a higher beam current, more ions can be generated by ionization; however,  stronger beam-ion interaction also enlarges the ions loss rate conversely. Without the synchrotron radiation damping,  the accumulated ion number  does not reach an equilibrium state after 10 thousand turns. More ions are accumulated since a larger beam action leading to a larger average distance between new generated ions and the centroids of coming bunches.

Fig.~\ref{fig:FigIonDenProfile} shows  the ion density profile variation at the 8000th, 8400th and 8900th turns, when the ``quasi-equilibrium" is arrived in the ``weak-strong" simulation with synchrotron radiation damping. The centre of the density profile shows an oscillation due to the beam-ion interaction and the oscillation amplitude is consistent with the maximum bunch oscillation in  Fig.~\ref{fig:WSFillingPattern0} and Fig.~\ref{fig:Syn_10mA_Mode_and_Oscillation}. The frequency of the ion density oscillation can be estimated by even filling pattern approximation \cite{9}. It is noticeable that the ion density profile deviates from the Gaussian type with a higher density peak in the center as discussed in Ref.~\cite{13}. In the future study, a self-consistent space charge solver is required for a better resolution.

\subsection{``Quasi-strong-strong" simulation}
\begin{figure*}[htb!]
    \centering
    \subfloat[\label{sfig:WSFillingPattern0BunchMode_1}]{
        \includegraphics[width=.45\linewidth]{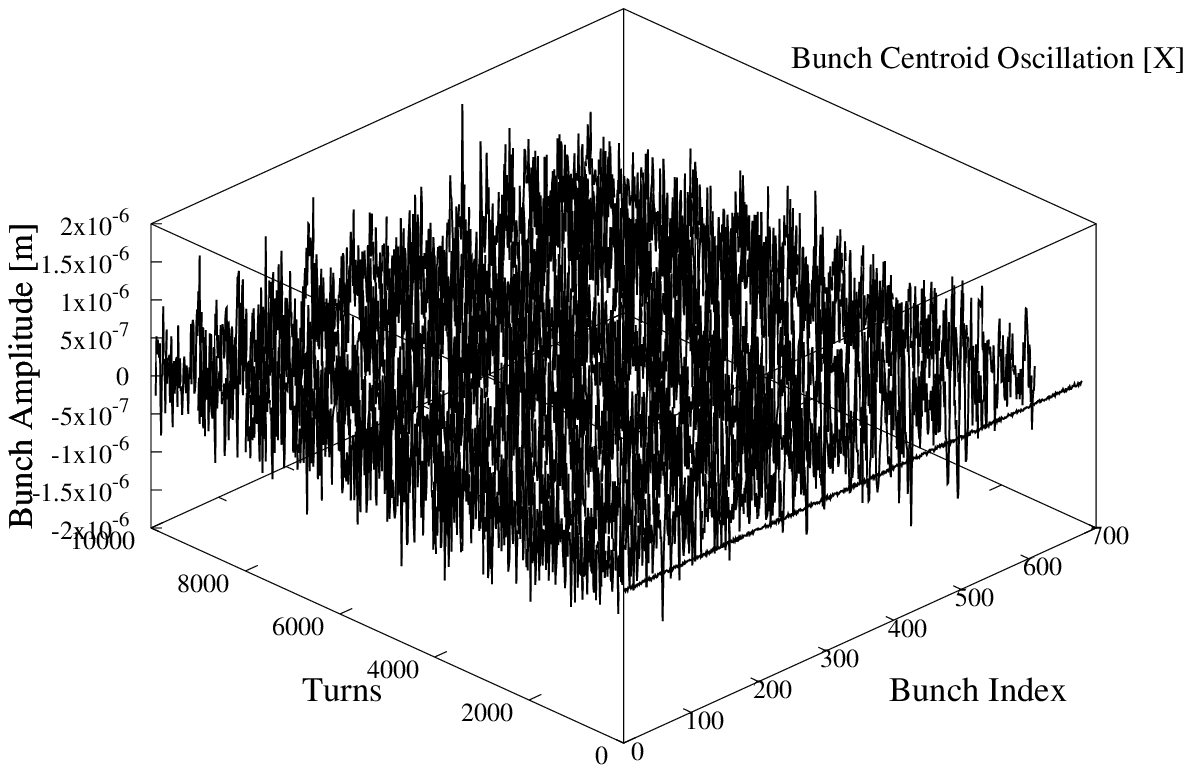}
    }
    \subfloat[\label{sfig:WSFillingPattern0BunchMode_2}]{
        \includegraphics[width=.45\linewidth]{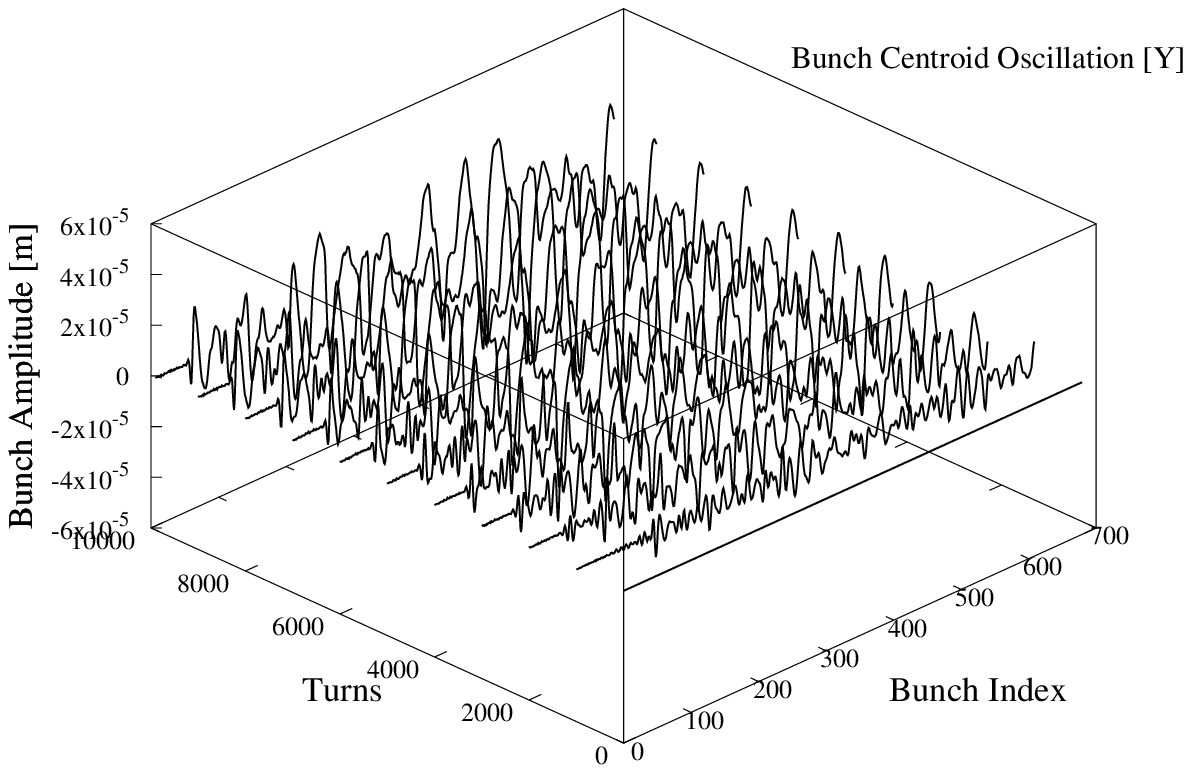}
    }
    
    \subfloat[\label{sfig:WSFillingPattern0BunchMode_3}]{
        \includegraphics[width=.45\linewidth]{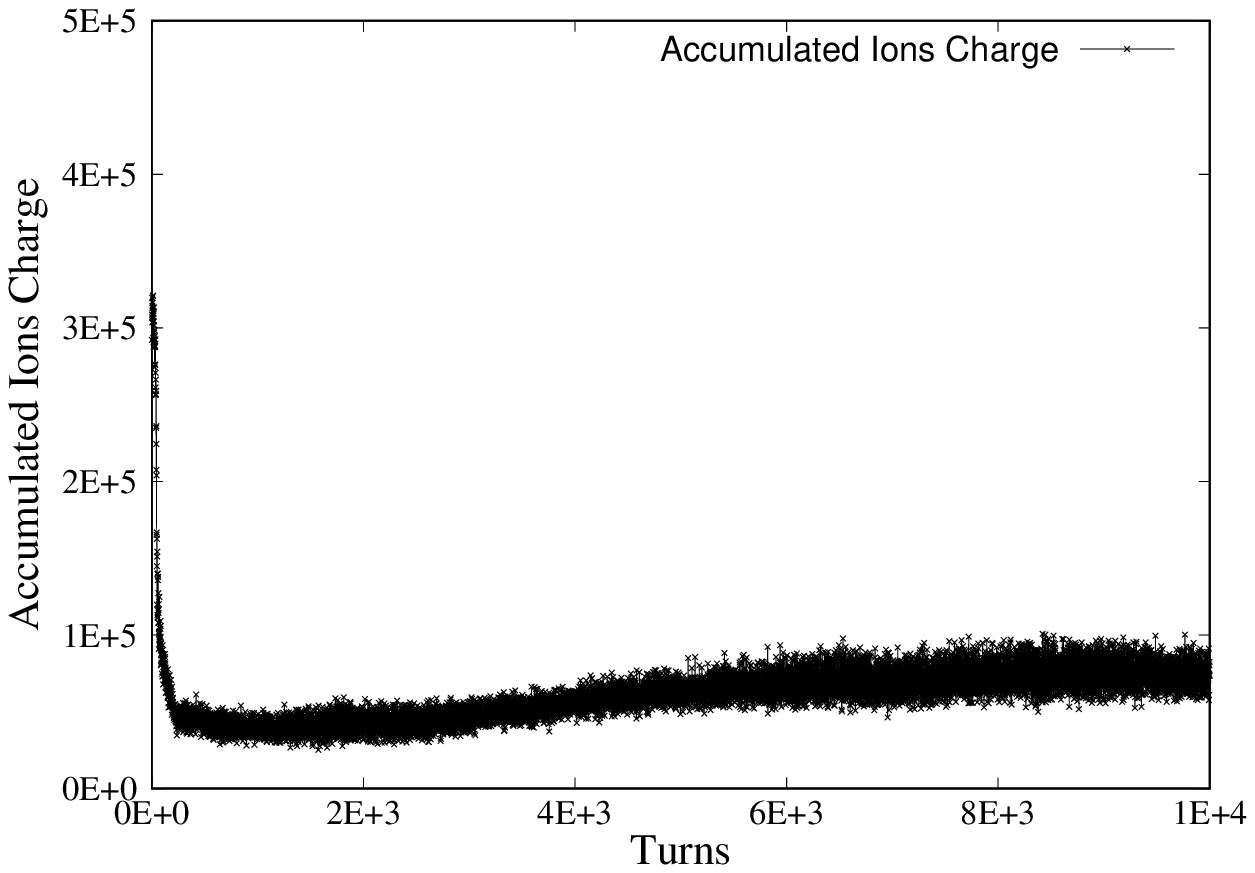}
    } 
    \subfloat[\label{sfig:WSFillingPattern0BunchMode_4}]{
        \includegraphics[width=.45\linewidth]{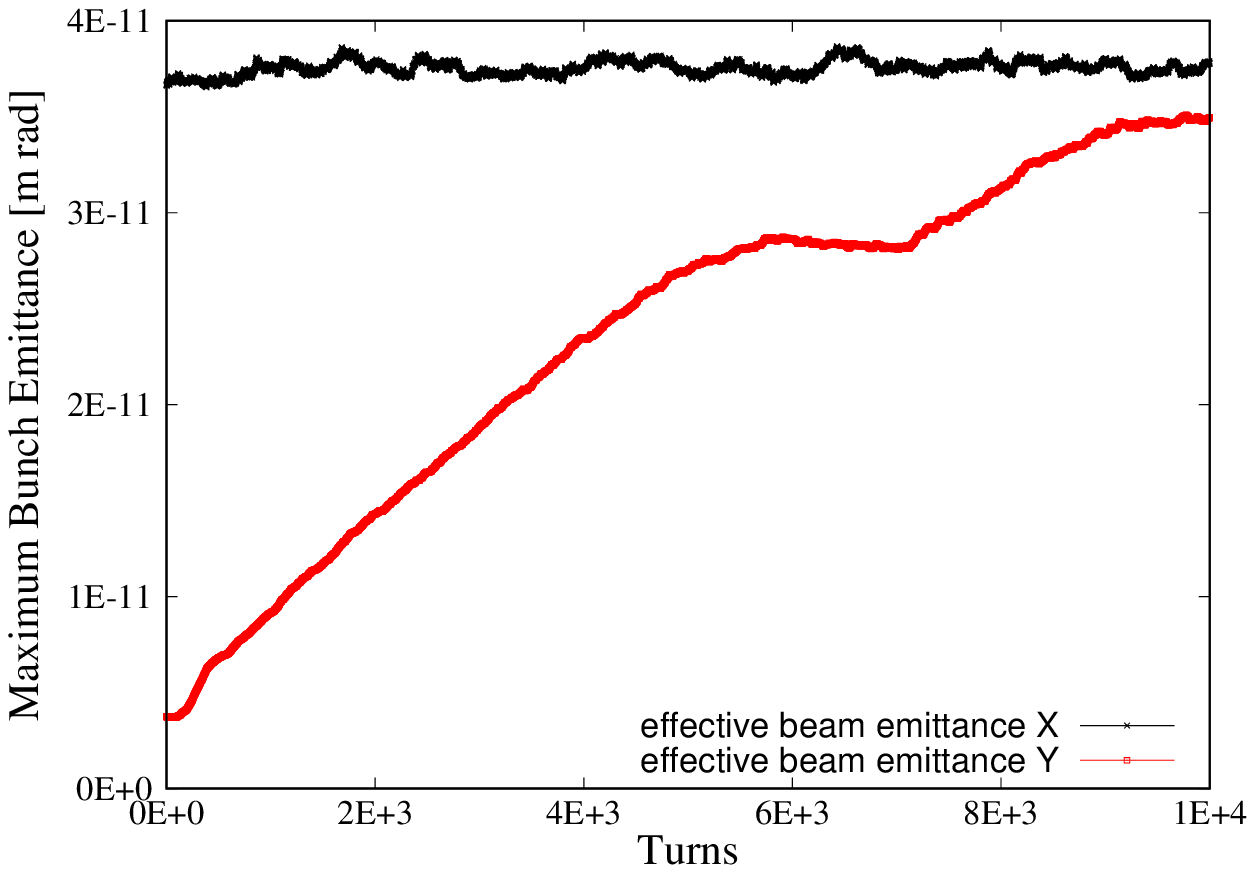}
    }
    \caption{Beam bunches oscillations in horizontal (a) and vertical (b) directions, accumulated ion charge (c) and the maximum bunch emittance references to the ideal orbit (d) as functions of passing turns;  The synchrotron radiation damping is taken into account. The beam current is 10 mA and the simulation results are given by ``quasi-strong-strong" model.}
    \label{fig:WSFillingPattern1BunchMode}
\end{figure*}

In the ``weak-strong" model,  the electron bunch is represented by one macro-particle and it can not give information on beam effective emittance growth. In this subsection, the ``quasi-strong-strong" model is adopted to evaluate the effective beam emittance growth \footnote{Rms emittance reference to the idea orbits.}. Both the ions and electron bunches are represented by macro-particles. Still, the Bassetti-Erskine formula  Eq.~\ref{eq2.5} is applied to get the space charge force between ion and beam as explained in section II.

Fig.~\ref{fig:WSFillingPattern1BunchMode} shows the bunch oscillations in horizontal and vertical directions, accumulated ions and beam effective emittance growth in ``quasi-strong-strong" simulation when the synchrotron radiation damping is turned on.  The maximum bunch centroid oscillation is roughly 0.1 and 10 times rms beam size in $x$ and $y$ direction respectively;  The accumulated ions charge gets smaller when the ``quasi-equilibrium" is reached. Fig.~\ref{sfig:WSFillingPattern0BunchMode_4} gives the maximum effective bunch emittance (reference to the ideal orbit) revolution as function of tracking turns.  The effective bunch emittance  continuously increases in  vertical direction and beyonds the error budget. The horizontal beam emittance does not change too much even after 10 thousand turns evolution. In the following section IV, a bunch-by-bunch feedback system based on Finite Impulse filters (FIR) filter techniques will be introduced to mitigate the beam-ion instability and control  the beam emittance growth.

\section{Bunch-by-bunch Feedback and its influence on beam performance}
The bunch-by-bunch feedback based on the FIR filter is an effective way to cure the coupled bunched instability. It detects transverse or longitudinal centroid positions of beam bunches, processes the positions data to create  kicker signals, and adds transverse or longitudinal kicks to the beam particles to damp its oscillations. Noticeably, in this process, only centroid oscillations of passed bunches are used.  The  momentum changes of particles in the bunched to be kicked  are the same. Thus, the bunch-by-bunch feedback damping effect is in an average scene.  In general, the spectrum of coupled bunch mode spectrum is 
\begin{eqnarray}\label{eq3.0}
Spectrum \propto M\omega_0\delta(\omega-\omega_{\beta}-pM\omega_0-\mu\omega_0) ,
\end{eqnarray}
where $M$ is the harmonic number, $p$ is integer, $\mu$ is the mode index and $\omega_0$ is the revolution frequency \cite{20}. The well designed FIR filter, with Direct Current (DC) rejection, has zero amplitude response at arbitrary harmonics of the revolution frequency $\omega_0$.  Clearly, by adopting the FIR filter, only the fraction of betatron oscillation frequency over revolution frequency $\omega_{\beta}/\omega_{0}$ matters.  That is the reason why the unstable coupled bunch mode can be damped.

Eq.~\ref{eq3.1} is the general form of a FIR filter
\begin{eqnarray}\label{eq3.1}
\Theta_n= \sum_{k=0}^{N}a_k x_{n-k},
\end{eqnarray}
where $a_k$ represents the filter coefficient, $x_{n-k}$ and $\Theta_n$ are the input and output of the filter, corresponding to beam position data at the $(n-k)$th turn and kick strength on the beam at the $n$th turn. The number of the input data $N+1$ is defined as taps. Following the approaches shown in Ref.~\cite{21}, the time domain least square fitting (TDLSF) method is used to get the filters  coefficients $a_k$.

\begin{figure*}[htb!]
    \centering
    \subfloat[\label{sfig:PhaseGainFilter9Tap0}]{
        \includegraphics[width=.3\linewidth]{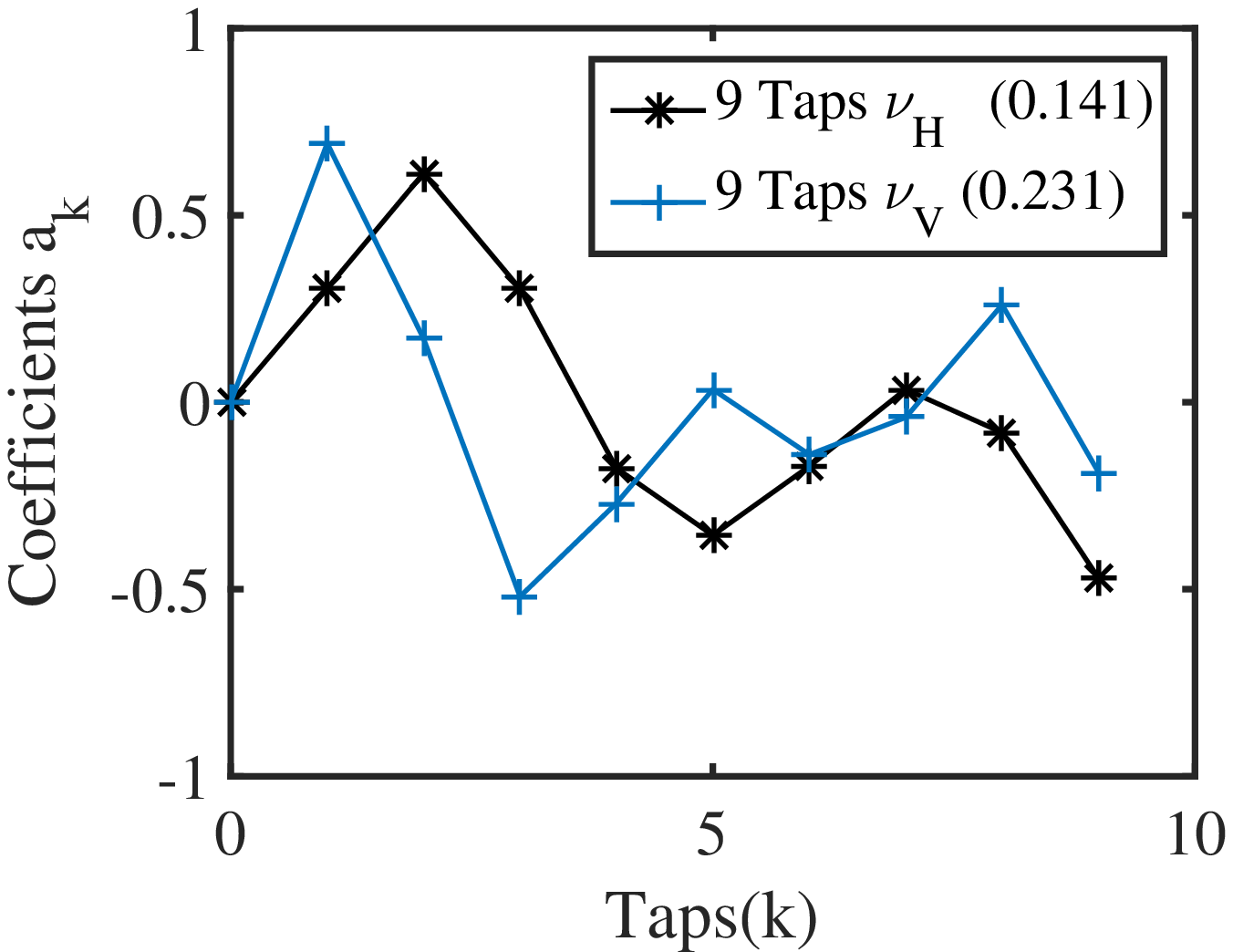}
    }
    \subfloat[\label{sfig:PhaseGainFilter9Tap1}]{
        \includegraphics[width=.3\linewidth]{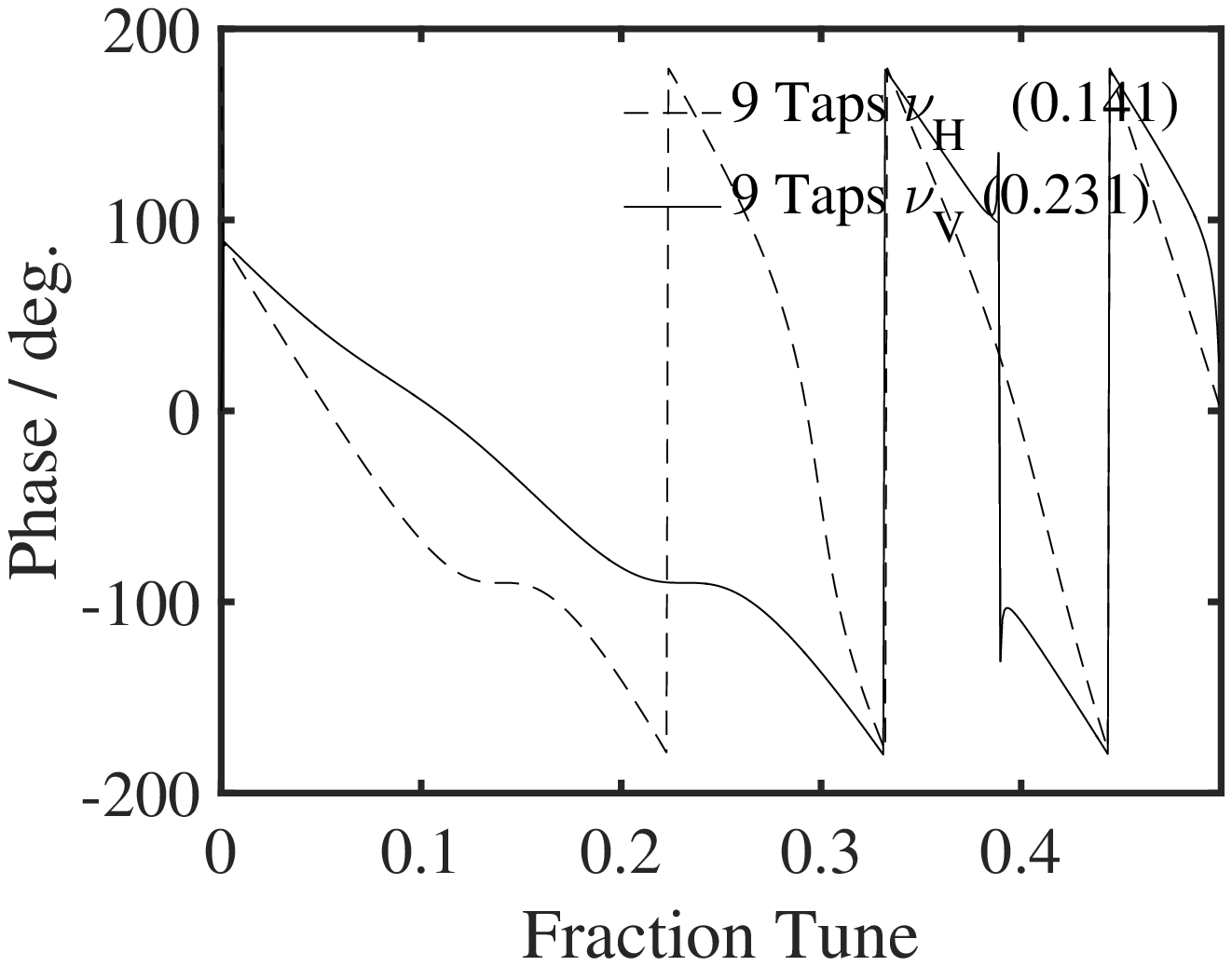}
    }
    \subfloat[\label{sfig:PhaseGainFilter9Tap2}]{
        \includegraphics[width=.3\linewidth]{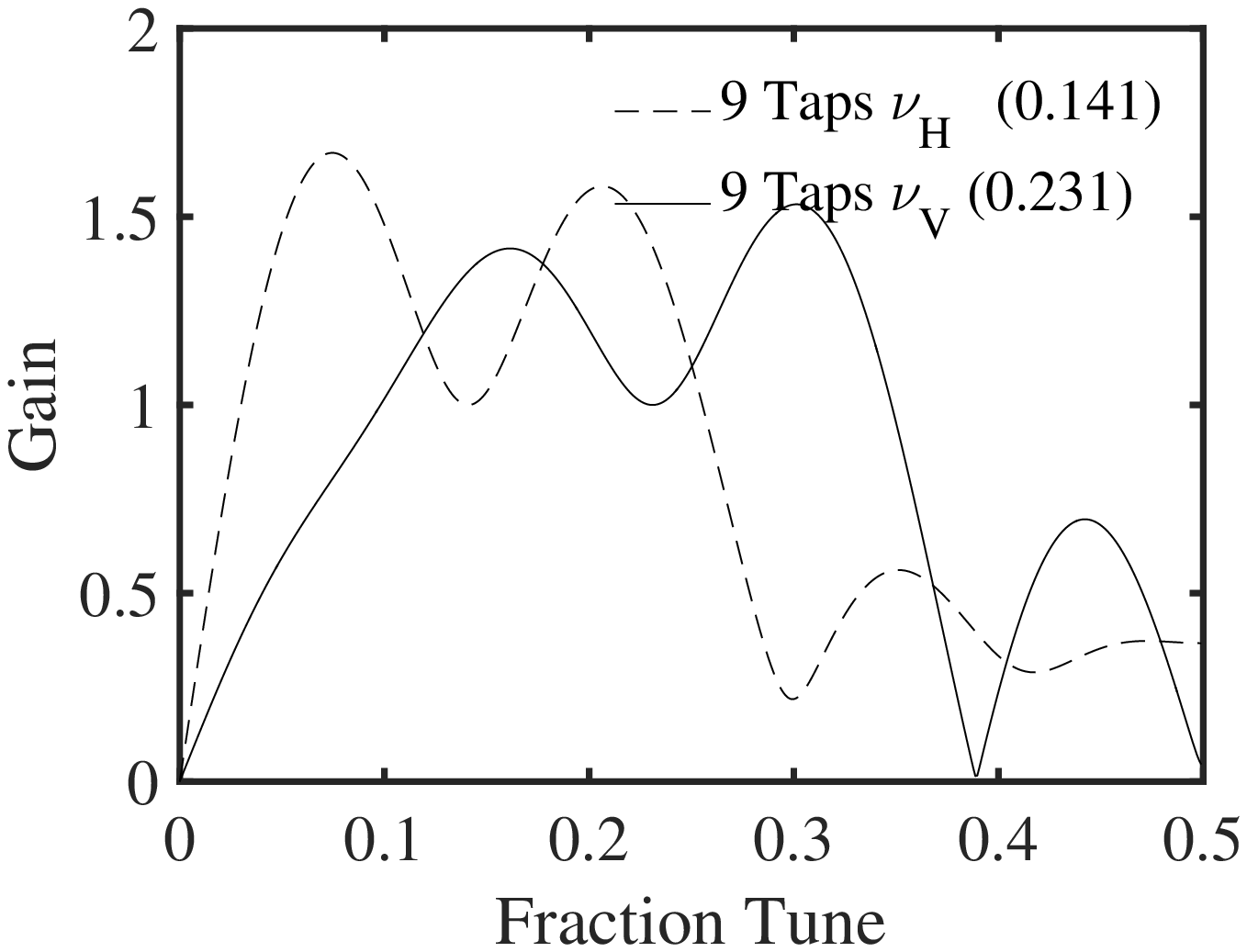}
    }
    \caption{FIR filter coefficients (a), frequency response of phase (b) and gain (amplitude) (c) of the 9-taps filters
    used in a bunch-by-bunch feedback system. The horizontal and vertical target tunes are 0.141 and 0.231.}
    \label{fig:PhaseGainFilter9Tap}
\end{figure*}

With the condition of one turn delay $a_0=0$, two 9-taps FIR filters are designed. For clarity, the pickup and kicker are assumed to be located at the same place with zero dispersion, which means the phase responses at target tunes are -90 degree. Fig.~\ref{fig:PhaseGainFilter9Tap} shows the filter coefficients, the phase and amplitude responses as function of fraction tune in horizontal and vertical directions. The first derivation of the phase response curves at target tunes are designed to be zero to enlarge the phase error tolerance. The stable working region corresponds to the  feedback phase response curve between  (-180,0) degrees. The gain at the target tune is local minimum and normalized to 1. The DC components which are caused by the closed orbit distortions, unequal bunch signal shapes from pickup electrodes and reflection at cable connections are rejected. 

The shortest damping time $\tau_{FB}$ of the feedback can be approximated \cite{14, 21} by
\begin{eqnarray}\label{eq3.7}
\frac{1}{\tau_{FB}} = \frac{f_{r} \sqrt{\beta_p \beta_k}}{2 h E/e } \frac{\sqrt{2P_{max} R_k} }{\Delta X_{max}} ,
\end{eqnarray}
where $f_{r}$ is revolution frequency, $h$ is the harmonic number, $\beta_p$ and $\beta_k$ are the betatron function at pickup and kicker, $E$ is the beam energy, $e$ is the electron unit charge, $\Delta X_{max}$ is the maximum bunch oscillation that the feedback can suppress, $R_{k}$ is the kicker shut impedance and $P_{max}$ is the maximum available kicker power. In HEPS, the kicker shut impedance is  $R_{k} = 123$ K$\Omega$ at the location with $\beta_p=\beta_k=5$m, assuming the power limit on kicker is $P_{max}=1$ KW and $\Delta X_{max}=0.1$ mm, the shortest damping time of the feedback can supply is roughly 60 $\mu s$.

\begin{figure*}[htb!]
    \centering
    \subfloat[\label{sfig:WSFillingPattern0BunchModeFBDamping_1}]{
        \includegraphics[width=.45\linewidth]{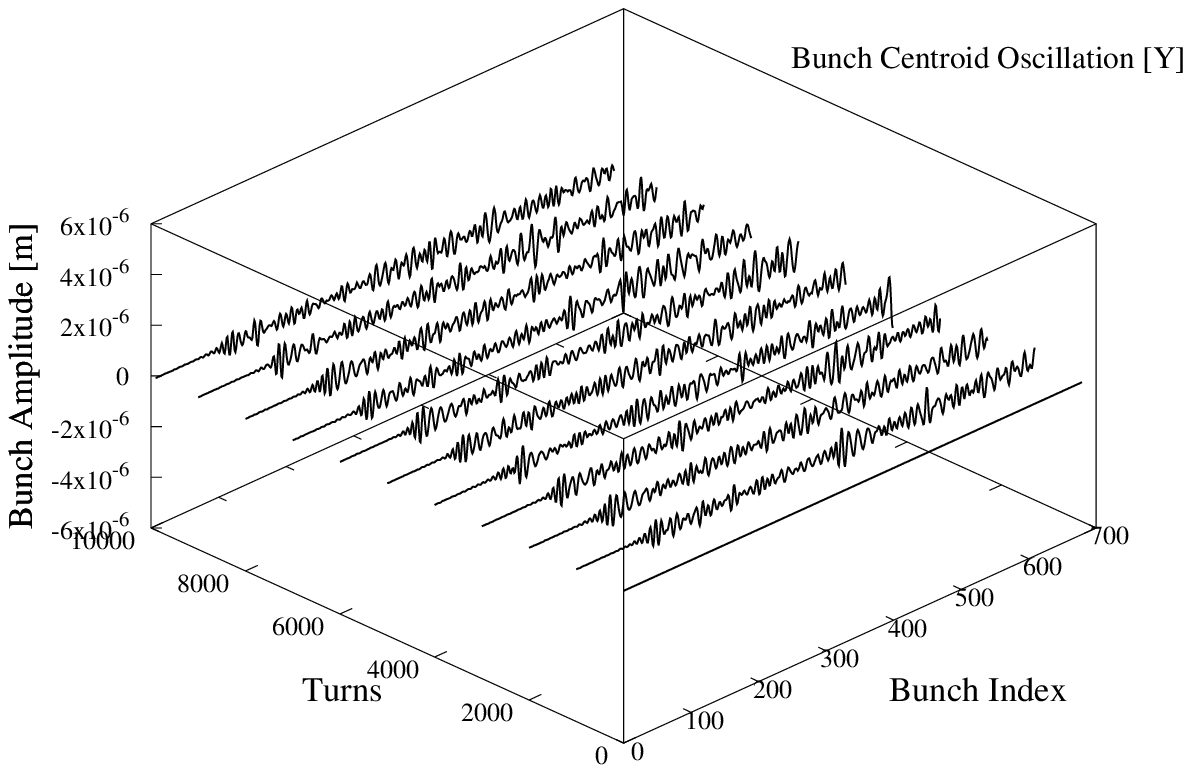}
    }
    \subfloat[\label{sfig:WSFillingPattern0BunchModeFBDamping_2}]{
        \includegraphics[width=.45\linewidth]{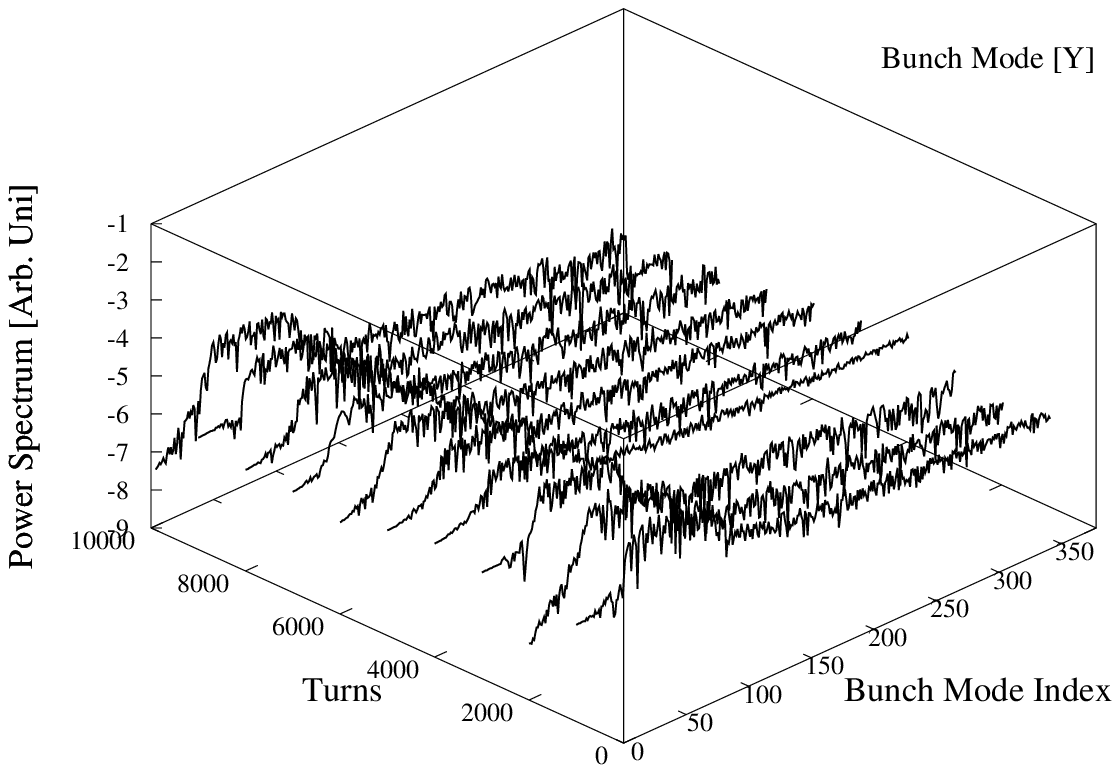}
    }
    \caption{Beam bunches oscillations in y (left) and its frequency spectrum (right) as function of passing turns,  both the synchrotron radiation damping  and bunch-by-bunch FIR feedback system are taken into account. The beam current is 10 mA and the ``weak-strong" model is used in simulation.}
    \label{fig:WSFillingPattern0BunchModeFBDamping}
\end{figure*}

In the code, the beam momentum change by the bunch-by-bunch feedback at the $n$th turn is modeled as 
\begin{eqnarray}\label{eq3.5}
\Theta_{x,n}= K_x \sum_{k=0}^{N}a_{k,x} x_{n-k}, \nonumber  \\
\Theta_{y,n}= K_y \sum_{k=0}^{N}a_{k,y} y_{n-k},
\end{eqnarray}
here $x_{n-k}$ and $y_{n-k}$ are the beam centroids  of the $k$th previous turn at the pickup. The  beam motion transfer function in one turn including feedback is 
\begin{eqnarray}\label{eq3.6}
 { 
 \left[
 \left( 
 \begin{array}{c} 
 x_{n+1}  \\ 
 x'_{n+1}  \\
  y_{n+1}  \\ 
 y'_{n+1}  
 \end{array} 
 \right)
 \right ]}
 =M_0{ 
 \left[
 \left(  
 \begin{array}{cc} 
 x_{n}  \\ 
 x'_{n}  \\
  y_{n}  \\ 
 y'_{n}  
 \end{array}
  \right)
 +  
 \left(  
 \begin{array}{cc} 
 0  \\ 
 \Theta_{x,n} \\
 0  \\ 
 \Theta_{y,n} \\ 
 \end{array}
   \right)
 \right ]},
\end{eqnarray}
 where $M_0$ is the one turn matrix at the  kicker.

\begin{figure*}[htb!]
    \centering
    \subfloat[\label{sfig:WSFillingPattern0ComFBandNoFB_1}]{
        \includegraphics[width=.3\linewidth]{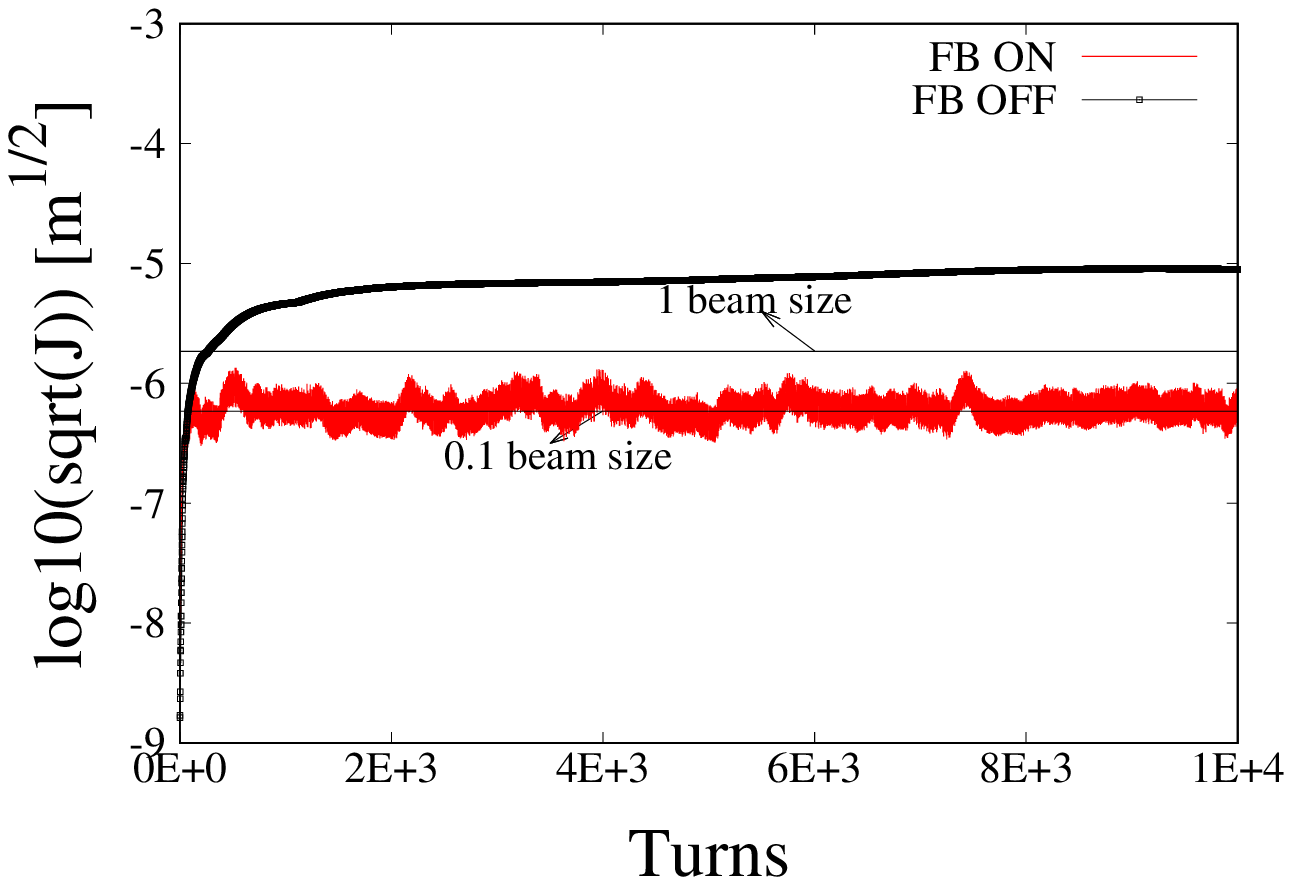}
    }
    \subfloat[\label{sfig:WSFillingPattern0ComFBandNoFB_3}]{
        \includegraphics[width=.3\linewidth]{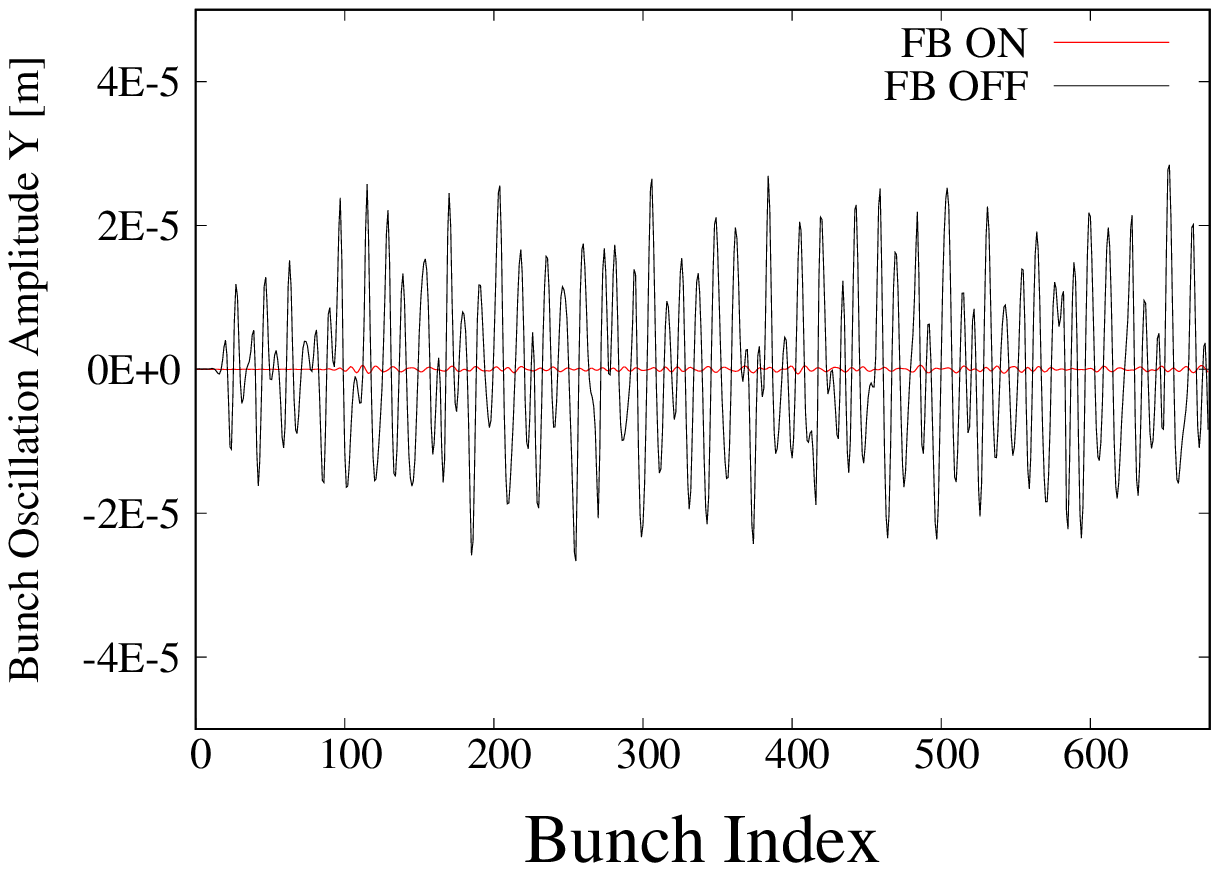}
    }
    \subfloat[\label{sfig:WSFillingPattern0ComFBandNoFB_4}]{
        \includegraphics[width=.3\linewidth]{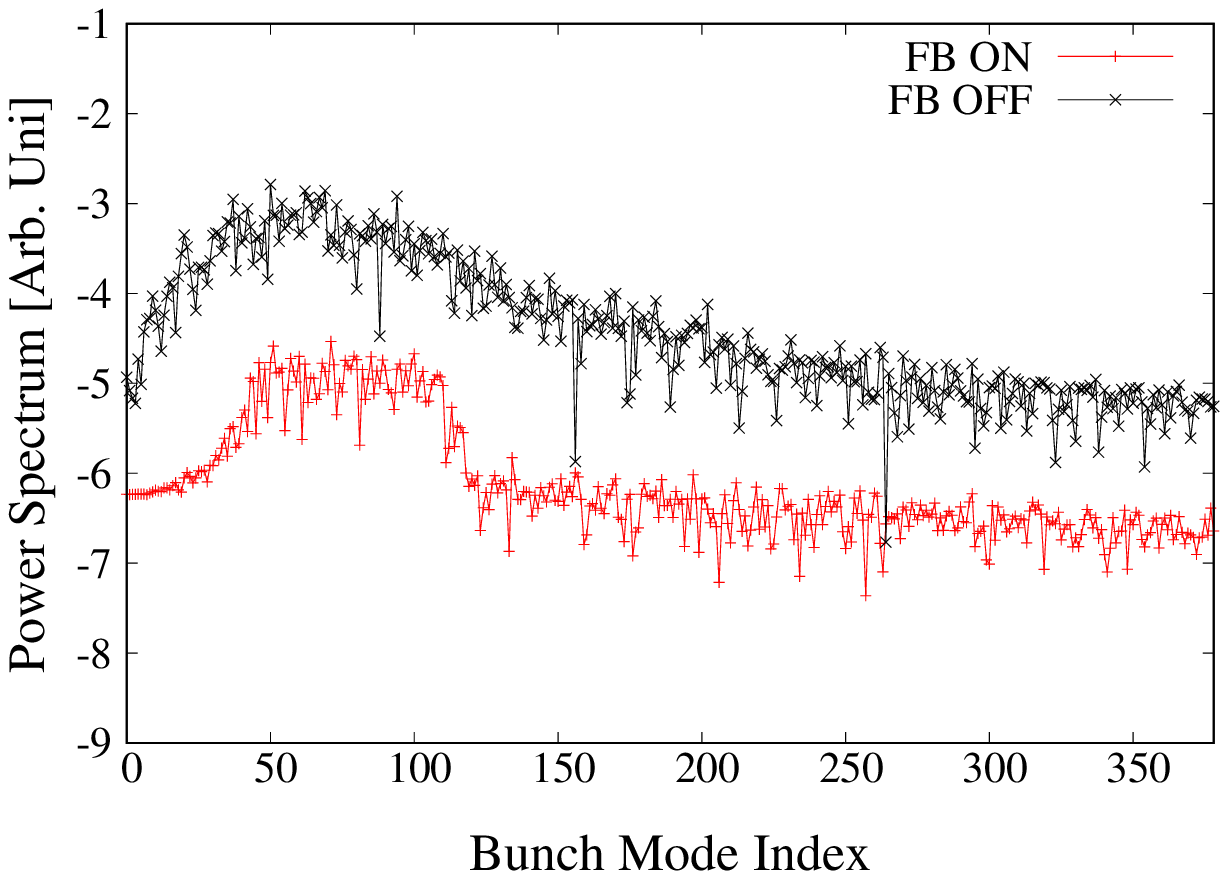}
    }
    \caption{The red and black curve show the maximum beam actions (a), the beam bunches oscillations (b) and the related coupled bunch modes power spectrum (c) in vertical (y) plane with and without bunch-by-bunch feedback at the 5000th turns.   The results are obtained from the ``weak-strong" model taking the synchrotron radiation damping into account.}
    \label{fig:WSFillingPattern0ComFBandNoFB}
\end{figure*}

\begin{figure*}[htb!]
    \centering
    \subfloat[\label{sfig:SS_SYN_FB_10mA_2}]{
        \includegraphics[width=.45\linewidth]{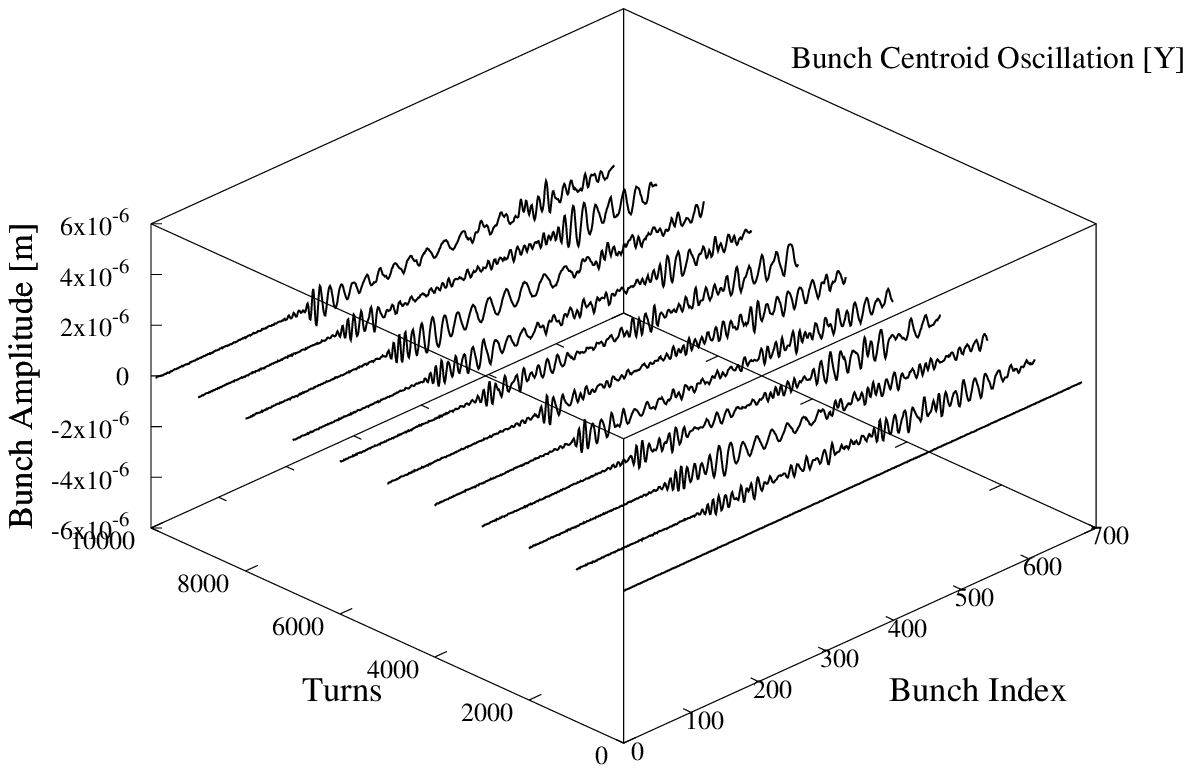}
    }
        \subfloat[\label{sfig:SS_SYN_FB_10mA_1}]{
        \includegraphics[width=.45\linewidth]{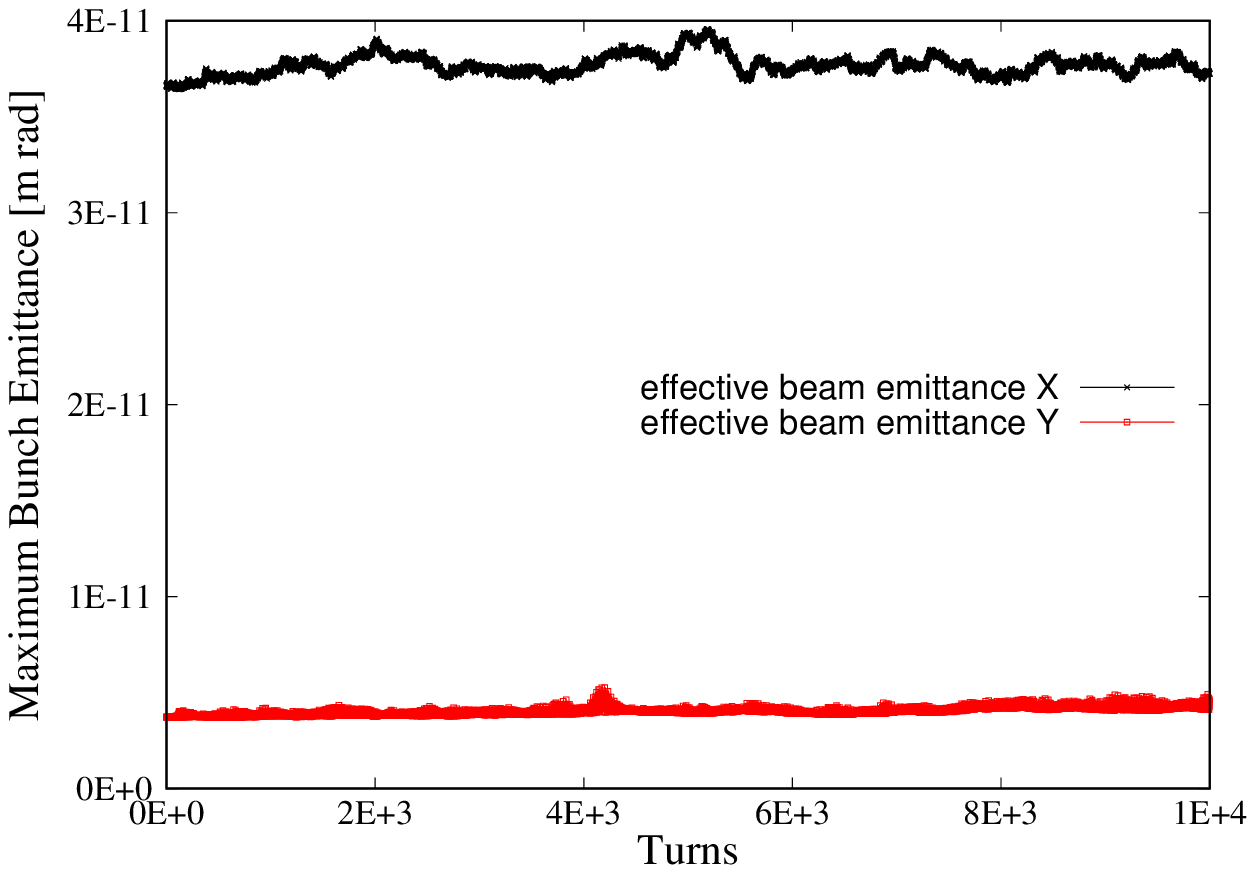}
    }
    \caption{Beam bunches oscillations (left)  and the effective rms emittance (right) in  vertical plane (y)  as function of passing turns, both the synchrotron radiation damping  and bunch-by-bunch FIR feedback system are taken into account. The results are obtained from the ``quasi-strong-strong" model.}
    \label{fig:SS_SYN_FB_10mA}
\end{figure*}

Fig.~\ref{fig:WSFillingPattern0BunchModeFBDamping} is calculated with the same parameters as in Fig.~\ref{fig:Syn_10mA_Mode_and_Oscillation} by the ``weak-strong" model expect that the bunch-by-bunch feedback is turned on. Clearly, the bunch centroid oscillation amplitudes are effectively damped down at least to one order of magnitude smaller compared with the case without bunch-by-bunch feedback. Comparison at the 5000th turn with and without feedback is explicitly shown in Fig.~\ref{fig:WSFillingPattern0ComFBandNoFB}. When the bunch-by-bunch feedback is turned on, the maximum bunch action is well maintained around 0.1 rms beam size, Fig.~\ref{sfig:WSFillingPattern0ComFBandNoFB_1}; the bunch oscillations due to beam-ion interaction are well eliminated, Fig.~\ref{sfig:WSFillingPattern0ComFBandNoFB_3}; the power spectrum of of bunch oscillations is roughly one order of magnitude smaller Fig.~\ref{sfig:WSFillingPattern0ComFBandNoFB_4}. The maximum unstable bunch mode does not shift since the intrinsic beam-ion interaction is not violated.

The simulation results given by the ``quasi-strong-strong" model are shown in Fig.~\ref{fig:SS_SYN_FB_10mA}. The bunch centroid oscillation amplitudes are well damped down within $\pm$0.1 rms beam size by the bunch-by-bunch feedback system. The effective rms beam emittance are well maintained and only show a tiny increasement after 10 thousand turns. Generally, the effective rms emittance growth can be attributed to two effects. The first one is the coherent bunch centroid oscillation and the second one is the bunch phase space filamention due to the nonlinear Coulomb force. Here, it is stated that, in beam-ion interaction, the coherent bunch centroid oscillation is the main contributor  to the effective emittance growth. The effective rms emittance growth from nonlinear space charge is almost inessential.

\section{Conclusions}
In this paper, we  have discussed the beam-ion instability and its mitigation by the bunch-by-bunch feedback system. To study the beam-ion interaction consistently, a simulation code is developed including modules of  ionization, beam-ion interaction, synchrotron radiation damping, quantum excitation and bunch-by-bunch feedback. As an example, the lattice parameters of HEPS are adopted to show the influence of the beam-ion effect. It is found that in  high intensity and ultra-low emittance rings, the beam-ion instability is not a dominant mechanism when the beam current goes  high enough. This is  because the ions generated are over-focused, get lost between the bunch gaps,  and hardly disturb the beam. The beam-ion interaction significantly impacts the beam performance only when ions can be extensively accumulated in certain current region. If the beam-ion instability is too strong to be suppressed by the synchrotron radiation damping, the bunch-by-bunch feedback based on FIR filter technique is proved to be an effective approach to rescue the beam performance.

Still, studies in this paper can be improved to further steps in the future. One point is adding a real self-consistent PIC solver into the code.  It is worthy to ensure a real self-consistent process especially when applying the results of  codes as guidance for real accelerators. Another point is related to  the impedance. As known, an abundant of instabilities can be induced by the impedance. It will be very important to understand how the  beam-ion interaction and the impedance interplay with each other. In the future, modules to deal with the impedance and wakefield will be developed to study the complex beam dynamics system and then the ability of feedback system will be re-evaluated.  

\begin{acknowledgments}
This work is supported by the Key Research Program of Frontier Sciences CAS (QYZDJ-SSW-SLH001) and NSFC (11775239, 11805217). Special thanks to professor Takeshi Nakamura in Synchrotron Radiation Research Institute in Japan for his helpful discussion on FIR filters.
\end{acknowledgments}

\bibliography{reference}
\end{document}